\documentclass[aps, preprint, nofootinbib,preprintnumbers,eqsecnum,superscriptaddress,sort]{revtex4}

\usepackage{color}

\usepackage[
      colorlinks=true,
      linkcolor=blue,
      urlcolor=blue,
      filecolor=black,
      citecolor=red,
      pdfstartview=FitV,
      pdftitle={},
        pdfauthor={Oscar Dias, Gary Horowitz, Nabil Iqbal, Jorge Santos},
        pdfsubject={},
        pdfkeywords={},
        pdfpagemode=None,
        bookmarksopen=true
      ]{hyperref}

\usepackage[normalem]{ulem}
\usepackage{amsmath}
\usepackage{enumerate}
\usepackage{amsfonts}
\usepackage{yfonts}

\usepackage{subfigure}
\usepackage{psfrag}

\usepackage{epsfig}
\usepackage[latin1]{inputenc}
\usepackage{float}
\usepackage{graphicx}
\usepackage{cancel}
\usepackage{mathrsfs}
\usepackage{amssymb}
\usepackage{amsfonts}
\usepackage{amsmath}
\usepackage{slashed}

\usepackage{graphicx}
\usepackage{bm}

\def\({\left(}
\def\){\right)}
\def\[{\left[}
\def\]{\right]}
\def\<{\langle}
\def\>{\rangle}

\def\CO{{\cal O}}

\newcommand\p{\ensuremath{\partial}}

\newcommand{\dd}{\mathrm{d}}

\newcommand{\be}{\begin{equation}}
\newcommand{\ee}{\end{equation}}
\newcommand{\bea}{\begin{eqnarray}}
\newcommand{\eea}{\end{eqnarray}}
\newcommand{\bwt}{\begin{widetext}}
\newcommand{\ewt}{\end{widetext}}

\newcommand{\bi}{\begin{itemize}}
\newcommand{\ei}{\end{itemize}}
\newcommand{\ben}{\begin{enumerate}}
\newcommand{\een}{\end{enumerate}}
\newcommand{\bca}{\begin{cases}}
\newcommand{\eca}{\end{cases}}
\newcommand{\bln}{\begin{align}}
\newcommand{\eln}{\end{align}}
\newcommand{\bst}{\begin{split}}
\newcommand{\est}{\end{split}}

\newcommand\ep{\epsilon}

\newcommand\lam{\lambda}
\newcommand\Lam{\Lambda}
\newcommand\om{\omega}

\newcommand\ga{{\ensuremath{{\gamma}}}}

\def\th{{\theta}}

\def\le{\left}
\def\ri{\right}

\newcommand\sO{{\ensuremath{{\mathcal O}}}}

\newcommand{\ka}{{\kappa}}

\pdfoutput=1

\begin{document}

\title {Vortices in holographic superfluids and superconductors as conformal defects}

\preprint{NSF-KITP-13-251}

\author{\'Oscar J. C. Dias}
\email{oscar.dias@ist.utl.pt}
\affiliation{Center for Mathematical Analysis, Geometry, \& Dynamical Systems, \\
Departamento de Matem\'atica and LARSyS, Instituto Superior T\'ecnico, 1049-001 Lisboa, Portugal}
\affiliation{Institut de Physique Theorique, CEA Saclay, CNRS URA 2306, F-91191 Gif-sur-Yvette, France}

\author{Gary T. Horowitz}
\email{gary@physics.ucsb.edu}
\affiliation{Department of Physics, UCSB, Santa Barbara, CA 93106}

\author{Nabil Iqbal}
\email{niqbal@kitp.ucsb.edu}
\affiliation{Kavli Institute for Theoretical Physics, UCSB, Santa Barbara CA 93106 }

\author{Jorge E. Santos}
\email{jss55@stanford.edu}
\affiliation{Department of Physics, Stanford University, Stanford, CA 94305-4060, U.S.A.}
\affiliation{Department of Applied Mathematics and Theoretical Physics, University of Cambridge, Wilberforce Road, Cambridge CB3 0WA, UK \\ \vspace{1 cm}\,}

\begin{abstract}
We present a detailed study of a single vortex in a holographic symmetry breaking phase. At low energies the system flows to an nontrivial conformal fixed point. Novel vortex physics arises from the interaction of these gapless degrees of freedom with the vortex: at low energies the vortex may be understood as a conformal defect in this low energy theory. Defect conformal symmetry allows the construction of a simple infrared geometry describing a new kind of extremal horizon: a Poincar\'e horizon with a small bubble of magnetic Reissner-Nordstr\"{o}m horizon inside it that carries a single unit of magnetic flux and a finite amount of entropy even at zero temperature. We also construct the full geometry describing the vortex at finite temperature in a UV complete theory. We study both superfluid and superconducting boundary conditions and calculate thermodynamic properties of the vortex. A study of vortex stability reveals that the dual superconductor can be Type I or Type II, depending on the charge of the condensed scalar. Finally, we study forces on a moving vortex at finite temperature from the point of view of defect conformal symmetry and show that these forces can be expressed in terms of Kubo formulas of defect CFT operators. 
\end{abstract}

\today

\maketitle

\tableofcontents


\section{Introduction}

Holography has opened a new window in the study of strongly correlated states of matter. This approach is particularly useful when dealing with many interacting gapless degrees of freedom, which is precisely where traditional field-theoretical methods fail yet the dual gravitational description is the most tractable. 

One phase of holographic matter that has been extensively studied is the holographic superfluid or superconductor \cite{Gubser:2008px,Hartnoll:2008vx,Hartnoll:2008kx}. The ground state of these holographic phases is very different from those of a conventional field-theoretical Bose superfluid or superconductor of the sort found in textbooks. A conventional superfluid contains very few low-energy excitations: at zero temperature, there need only be a single gapless Goldstone mode associated with the breaking of a spontaneous symmetry. A conventional superconductor will not possess even this mode: it will be eaten by the dynamical photon. However, a typical {\it holographic} superfluid or superconductor 
 will often possess many
 gapless degrees of freedom, as can be seen geometrically in its gravity dual from the existence of a
 black hole horizon at low temperature. This horizon has an entropy scaling like a power of $T$ with  a large coefficient
 \cite{Horowitz:2009ij,Gubser:2009cg}. These degrees of freedom have nothing to do with any Goldstone mode, and in many cases there is an emergent scaling or even conformal symmetry controlling this low-energy physics. 

The robust coexistence of these gapless modes together with symmetry-breaking order is somewhat novel from a field-theoretical point of view. In this paper we will study a consequence of this cohabitation by probing the IR structure with an excitation that all such phases possess: a vortex. Previous studies of holographic superfluid and superconducting vortices include \cite{Albash:2009iq,Montull:2009fe,Keranen:2009re,Maeda:2009vf,Domenech:2010nf,Bao:2013fda,Bao:2013ixa}. The basic idea is well-understood: the vortex becomes a cosmic string in the bulk, carrying magnetic flux down to the bulk horizon. Many of the previous studies are in a probe limit, or take backreaction into account perturbatively, which is completely well-defined only at finite temperatures. Our treatment will improve on their work by going to zero temperature and thus truly studying infrared physics. This will require us to include backreaction and thus numerically construct a new class of black hole solutions, resulting in conceptually new ingredients, which we summarize below. 

\subsection{Motivation and summary of results}

We study the (3+1)-dimensional gravitational dual to a (2+1)-dimensional  superfluid or superconductor. In the UV, our system is a conformal field theory.  We consider the case where the zero temperature  bulk solution is a domain wall in the holographic direction: in the IR, the system exhibits an emergent AdS$_4$ region. This means that the infrared degrees of freedom have rearranged themselves into a {\it different} conformal field theory, which we refer to from now on as the {\it IR CFT}. Our goal is to study the interaction of the vortex with these new degrees of freedom. We will actually study the theory with both superfluid and superconducting (in which the boundary symmetry is gauged) boundary conditions, finding very different vortex physics, as expected. 

The study of this vortex is interesting from several different points of view. From a gravitational point of view, the vortex in the bulk is a cosmic string that carries a single unit of magnetic flux. The vortex line extends all the way from the conformal boundary to a new IR Poincar\'e horizon. The physics where this flux meets the degenerate horizon is nontrivial.\footnote{For a discussion of cosmic strings piercing a finite temperature horizon, see \cite{Achucarro:1995nu}.} The condensate vanishes at the core of the vortex, where the magnetic flux is focused: thus we expect to find a new kind of black hole horizon, containing a small bubble of extremal magnetic Reissner-Nordstr\"{o}m horizon, surrounded by a sea of superconducting Poincar\'e horizon. Associated with this piece of Reissner-Nordstr\"{o}m horizon is a finite $T = 0$ ``impurity'' entropy which may be associated with the presence of the vortex. One of the main results of this paper is an explicit construction of this new kind of horizon. 

This horizon structure has an elegant understanding from the field theory. The fact that the vortex extends into the horizon means that it interacts nontrivially with the IR degrees of freedom. There is a well-developed formalism to deal with such a situation, that of {\it defect CFT} \cite{Cardy:2004hm}, which deals with the interaction of heavy objects -- such as a single vortex -- with a gapless conformal field theory. Previous study of similar defects at critical points separating antiferromagnetic order from a paramagnetic phase includes \cite{MaxSubir,Sachdev24121999,PhysRevB.61.15152}. One concrete consequence is that there is a reduced conformal symmetry, corresponding to what remains if we remove the translation generators from the full conformal group. This residual conformal symmetry turns out to be enough to reduce the PDEs determining the full gravitational solution to a (relatively) simple set of ODEs that determines the physics of the deep infrared at zero temperature. 

Furthermore, various observables characterizing the vortex can be calculated in terms of operators living on the conformal defect. 
In conventional superfluids, the precise form of the forces acting on a superfluid vortex in motion can be a matter of considerable controversy. However, in a conformal superfluid of the sort described here, there are precise Kubo formulas for these forces, written in terms of correlators of operators localized on the defect. This feature is independent of our gravitational description, and we anticipate further applications. 

While the infrared physics of the system is elegant, we do not restrict ourselves to this limit. We also explicitly solve the partial differential equations corresponding to the vortex at a finite temperature, demonstrating that the IR features discussed above emerge from the full gravitational solution in the $T \to 0$ limit. We also discuss some less universal physics: a novel result concerns the stability of superconducting vortices, i.e. whether a vortex with 2 units of flux is unstable to dissociating into two vortices. This feature is correlated with whether the superconductor is Type I or Type II: interestingly, we find that depending on the charge of the scalar the superconductor may be Type I. (See \cite{Umeh:2009ea} for an earlier indication that holographic superconductors can be Type I.)

Finally, there is one further reason to study the physics of holographic vortices \cite{Bao:2013fda,Bao:2013ixa}. It has recently been shown that magnetic monopole operators are likely to play an important role in the characterization of finite-density holographic matter \cite{Faulkner:2012gt,Sachdev:2012tj}. For example, let us imagine taking the bulk S-dual of a holographic superconductor. This is now a phase in which a {\it magnetically} charged scalar field has condensed in the bulk. The quanta of this field are magnetic monopoles. In this S-duality frame the bulk gauge field is {\it confined} (and not Higgsed), and it is now electric (and not magnetic) flux that is forced into tight flux tubes. Where these electric flux tubes intersect the boundary, they appear in the dual field theory as localized point charges: thus the dual field theory is one with a charge {\it gap}, and we are studying the holographic dual of an insulator.\footnote{This is qualitatively different from the holographic insulator one gets using the AdS  soliton. In that case all excitations are gapped due to a global property of the solution. Here, electric field is locally confined in the bulk.} Thus the vortex solutions that we study may be viewed through the lens of S-duality as also  determining the internal structure of the gapped charges that exist in a novel insulating phase. To avoid notational confusion we will not perform any further S-dualities in the bulk of this paper, but it may be helpful to keep this S-dual interpretation in mind, and we will return to it in the conclusion. 

This S-dual interpretation was one motivation for how we break the $U(1)$ symmetry. If we start, as usual, with nonzero chemical potential, then the S-dual description will have localized electric charges in a background magnetic field. Instead we work with zero chemical potential,
 but deform the theory by a relevant double-trace operator which triggers  a nonzero scalar condensate. Another motivation for this form of symmetry breaking is purely technical: there is one less bulk function to solve for. 

We conclude this section with a brief outline of the paper, explaining how the results mentioned above are organized. In the remainder of this introduction we discuss further the infrared physics of vortices. In Section \ref{sec:setup} we outline the gravitational setup and explain the homogeneous symmetry broken phase that we study. In Section \ref{sec:confdef} we elaborate on the interpretation of the vortex as a conformal defect and numerically construct the geometry that captures the infrared physics. In Section \ref{sec:pdenum} we turn to the solution of the full problem at all energy scales and explain the relevant numerical methods and boundary conditions required to solve the PDEs. In Section \ref{sec:results} we present the results from this analysis, including a detailed discussion of vortex stability and thermodynamics. Section \ref{sec:confforce} is somewhat different and does not require a gravitational description: here we point out that the forces on a moving vortex can be expressed in terms of Kubo formulas of defect-localized operators. We present a summary and outline some directions for future work in Section \ref{sec:concl}. The Appendix contains some technical details.

\subsection{Infrared physics of vortices}
We devote the rest of this introduction to an explanation of the low-energy structure of vortices in holographic superfluids and superconductors and how it differs from that in conventional superfluids and superconductors. A conventional superfluid has only a Goldstone mode at low energies, whose action is given by
\be
S = \rho_s\int d^3 x\;(\nabla \th)^2 \ .
\ee
A vortex configuration is simply one where the phase $\th$ winds around a point, i.e. if we denote the azimuthal coordinate by $\varphi$ we have $\th(\vec{x}) \sim n\varphi$ with the vortex charge $n \in \mathbb{Z}$. This description breaks down at the origin, where the condensate is forced to vanish. As we discuss in detail later, the winding in $\th$ results in an extended current flow and a logarithmic IR divergence of the energy of a single superfluid vortex.

Now this action resembles that of a massless scalar, and so one might imagine that there is a conformal structure associated with even ordinary superfluids at low energies. This is not quite correct. $\th$ is the phase of the scalar condensate, and is periodic, $\th \sim \th + 2\pi$. Thus it cannot have a scaling dimension, and in any spacetime dimension higher than $2$, Goldstone modes are not conformal. Indeed, in the action written above $\rho_s$ has mass dimension $1$, and thus provides a scale. Thus one expects that for problems where the compact nature of $\th$ is important (such as those involving vortices), the theory is empty below the scale $\rho_s$, as the Goldstone mode effectively decouples. 

For a conventional superconductor the effective action is different: here we have a dynamical gauge field $a$, and the coupling above is modified to:
\be
S_{SC} = \rho_s \int d^3x\;(\nabla \th - q a)^2\,.
\ee
The vortex configuration here is slightly different: we still have $\th \sim n\varphi$, but now the dynamical gauge field tracks this phase, so that far from the core we have $a_{\varphi} = \frac{n}{q}$. This cuts off the logarithmic divergence of the energy, making the vortex a localized excitation. This is related to the fact that the long-range Goldstone mode has been eaten by this gauge field by the familiar Higgs mechanism. Thus all excitations are gapped, and the situation is in some ways even simpler. 

This simple low-energy behavior is not the case for holographic vortices; due to the presence of other gapless modes, here we have a nontrivial conformal structure at arbitrarily low energies. However one could still ask whether the vortex necessarily {\it needs} to interact with this conformal structure. After all, the definition of the vortex is in terms of its interactions with the Goldstone mode, and so perhaps like the Goldstone mode above the vortex too could decouple from the low-energy dynamics. 

In a holographic system there is an interesting topological obstruction to such a decoupling. From the bulk point of view, for the vortex to decouple at low energies the vortex line must actually somehow end at some radial coordinate above the horizon. In the bulk we {\it always} have a dynamical gauge field, and thus the bulk vortex line carries magnetic flux. For it to end we thus must terminate it on a magnetic monopole in the bulk. 

\begin{figure}[ht]
\centering
\includegraphics[width=0.7\textwidth]{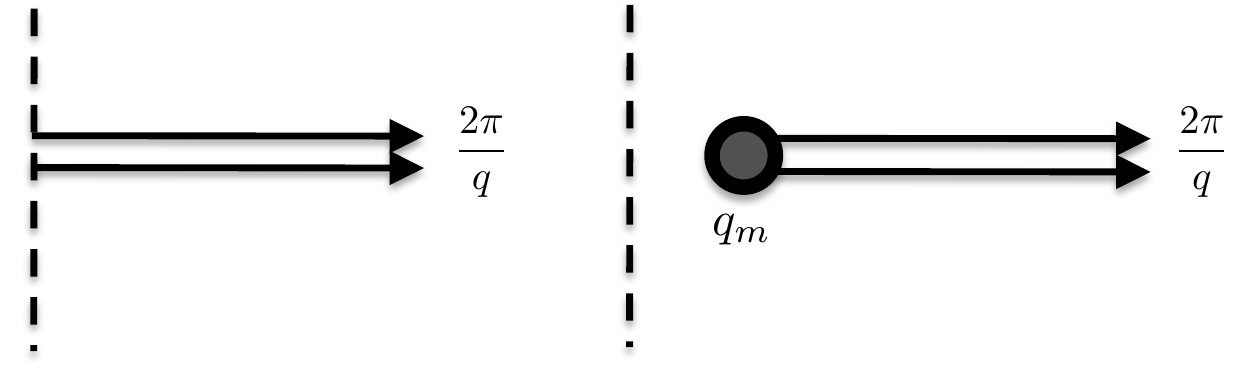}
\caption{Two different possibilities for infrared behavior of a holographic vortex. {\it Left}, vortex extends into horizon (dotted line) and never decouples. {\it Right}, vortex line terminated in bulk by magnetic monopole.}
\label{fig:monopole}
\end{figure}

This is not always possible. As the bulk $U(1)$ gauge group is compact, part of the definition of the theory is the specification of the smallest possible unit of electric charge $q_e$. The set of possible bulk magnetic monopole charges $q_m$ is determined by the Dirac quantization condition:
\be
q_e q_m = 2\pi \mathbb{Z} \ . \label{Dirac}
\ee 
Now the bulk magnetic flux carried by the vortex is $\frac{2\pi}{q}$, where $q$ is the charge of the condensed scalar field. $q$ is a multiple of the basic unit $q_e$. Now we see that if $q = q_e$ -- i.e. if we have condensed a scalar field with the smallest possible charge -- then we can terminate the vortex with a monopole, as shown in Fig.~\ref{fig:monopole}. Whether or not this actually happens depends on the dynamics (i.e. the balance between the bulk monopole mass and the tension in the string), but it is at least topologically possible. 

On the other hand, if  instead $q$ is some higher multiple $n q_e$, with $n > 1$, then the flux carried by the vortex is $1/n$ times the basic unit of magnetic flux, and the Dirac condition does not permit the existence of the fractionally charged magnetic monopole that would be required to terminate the line. Thus the line must extend to the horizon and cannot decouple from the conformal dynamics. From the field theory side, $q_e$ is the minimum charge quantum associated with the field theory Hilbert space. A vortex carrying flux $\frac{2\pi}{n q_e}$ will always have a nontrivial Aharonov-Bohm phase with elementary field theory quanta of charge $q_e$. Thus the field theory always knows of its existence and it cannot decouple. This line of reasoning appears to be an example of a general theme in applied holography: the value of the field theory charge quantum $q_e$ can manifest itself in bulk dynamics through the existence of magnetic monopoles \cite{Faulkner:2012gt,Sachdev:2012tj}. 

We now turn away from these general considerations to explicit computations in the bulk.

\section{\label{sec:setup}Setup of gravitational problem}

To describe a superconducting (or superfluid) vortex, we must first start with a homogeneous holographic superconductor (or superfluid). 
The simplest such theory  consists of gravity coupled to a Maxwell field and charged scalar, so we will work with the following action:
\begin{equation}
S= \frac{1}{16 \pi G_N}\int \dd^4 x\,\sqrt{-g}\left[R+\frac{6}{L^2}-\frac{1}{2}F_{ab}F^{ab}-2(D_a\Phi) (D^a\Phi)^{\dagger}  -2V(|\Phi|^2)\right],
\label{eq:action}
\end{equation}
where  $L$ is the AdS length scale, $F = dA$, and $D_a \Phi=\nabla_a \Phi -i\,q\,A_a \Phi$. The equations of motion read
\begin{subequations}
\begin{equation}
G_{ab}\equiv R_{ab}+\frac{3}{L^2}g_{ab}-\left[(D_a \Phi) (D_b \Phi)^\dagger+(D_b \Phi) (D_a \Phi)^\dagger+g_{ab}V(|\Phi|^2)+F_a^{\phantom{a}c}F_{bc}-\frac{g_{ab}}{4}F^{cd}F_{cd}\right]=0\,,
\label{eq:einsteinEOM}
\end{equation}
\begin{equation}
\nabla_a F^{ab}=i\,q\,\left[(D^b\Phi)\Phi^\dagger-(D^b\Phi)^\dagger \Phi \right],
\label{eq:maxwellEOM}
\end{equation}
\begin{equation}
g^{ab}D_aD_b \Phi-V'(|\Phi|^2)\Phi=0\,.
\label{eq:scalarEOM}
\end{equation}
\label{eqs:EOM}
\end{subequations}
It will often be convenient to use  $U(1)-$gauge invariant variables, which are defined  in terms of the  gauge field $A$ and complex scalar field $\Phi$ as
\begin{equation}
M = A-\frac{1}{q}\dd \tilde{\varphi},\quad\text{and}\quad \Psi = |\Phi|\,,
\end{equation}
where $\tilde{\varphi}$ is the phase of the complex scalar field $\Phi$.

\begin{figure}[ht]
\centering
\includegraphics[width=0.5\textwidth]{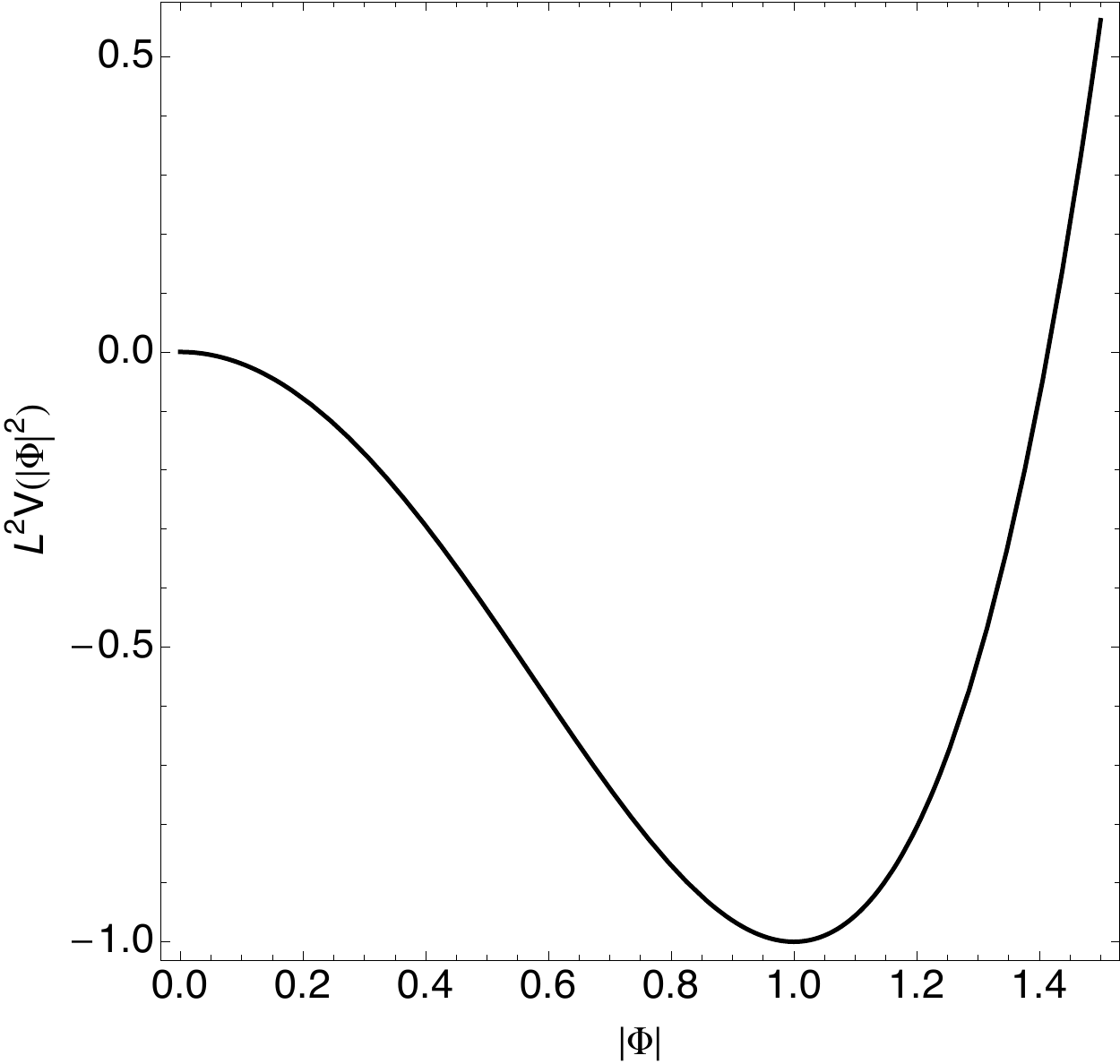}
\caption{Choice for the potential (\ref{eq:potential}), with $\mu^2 L^2 = -2$ and $V_0=-L^{-2}$.}
\label{fig:potential}
\end{figure}

We will choose our potential $V(|\Phi|^2)$ to be a standard Mexican hat potential, parametrized in the following way:
\begin{equation}
V(\eta)=\eta\,\mu^2\left(1-\frac{\eta\,\mu^2}{4\,V_0}\right)\,.
\label{eq:potential}
\end{equation}
This potential has two local extrema: one at $\eta=0$, where $V=0$ and another at $\eta = 2V_0/\mu^2$, where $V=V_0$. Furthermore, the mass of the complex scalar field at $\eta=0$ is given by $\mu^2$, whereas at $\eta = 2V_0/\mu^2$ we find an effective mass of $-\mu^4/(4\,V_0)$. Throughout the paper we will use $\mu^2 L^2 = -2$ and $V_0=-L^{-2}$, see Fig.~\ref{fig:potential}.

To describe vortices at finite temperature, we first need a homogeneous phase with a nonzero scalar field outside a black hole. One way to arrange this is to start with a charged black hole \cite{Hartnoll:2008kx}. However, the essential vortex physics that we would like to study does not require the complication of a nonzero background charge density. A different way to get the scalar field to condense is to add a double trace deformation in the boundary field theory \cite{Faulkner:2010gj}. This can cause nonzero scalar fields outside a neutral black hole as we now review.

We require that solutions asymptotically approach  AdS in Poincar\'e coordinates, i.e.
\begin{equation}
\dd s^2 = \frac{L^2}{z^2}\left(-\dd t^2+\dd R^2+R^2\dd \varphi^2+\dd z^2\right)\,.
\end{equation}
The asymptotic behavior of $\Phi$ is then
\be\label{eq:asympscalar}
\Phi = \alpha z + \beta z^2 + \cdots
\ee 
One has a choice of boundary conditions. For standard boundary conditions,  $\alpha = 0$, $\Phi$ is dual to a dimension two operator. For alternative boundary conditions, $\beta =0$, $\Phi$ is dual to a dimension one operator $\CO$. In this case, the double trace operator $\CO^\dagger \CO$ is relevant, so it is natural to add a coupling $-\ka \int d^3 x\;\CO^{\dagger} \CO$ to the dual field theory action. As explained in \cite{Witten:2001ua,Sever:2002fk}, the effect of adding such a term is to modify the boundary conditions in the bulk to become
\be\label{eq:double}
\beta = \kappa \alpha\,.
\ee
Positive $\kappa$ corresponds to adding $\CO^\dagger \CO$ to the dual field theory potential with a positive coefficient. This makes it harder for $\CO$ to condense. One might have thought that setting $\kappa < 0$ would destabilize the theory and there would be no ground state. However this is not the case. The full effective potential contains higher powers of $\CO$ which stabilize the theory. This has been shown by proving a bulk ``positive energy theorem" under the boundary condition $\beta = \kappa \alpha$ for $\Phi$ with $\kappa < 0$ \cite{Faulkner:2010fh}. 

For a given $\kappa < 0$,  the planar Schwarzschild solution (with $\Phi = 0$) is stable at high temperature, but becomes unstable to developing scalar hair at low temperature. The critical temperature is set by the only scale in the problem, $\ka$, and can be explicitly computed \cite{Faulkner:2010gj}:
 \be
T_c = {3\over 4\pi} {\Gamma(1/3)^3\over \Gamma(-1/3) \Gamma(2/3)^2} \,\kappa  \approx -0.62\, \kappa\,.
\ee
As $T \rightarrow 0$, we are deep in the condensed phase, and the value of the scalar field on the horizon approaches $|\Phi | = 1$. The horizon reduces to the Poincar\'e horizon of a new IR $AdS_4$ geometry. The $T=0$ solution thus interpolates between the UV $AdS_4$ with $\Phi=0$ and this new IR $AdS_4$ with $|\Phi| = 1$. Since we have chosen the minimum of $V$ to be $-1/L^2$, the effective cosmological constant in the deep infrared has increased from its UV value. This corresponds to a smaller effective AdS length, related to $L$ as $\tilde{L}^2= 3\,L^2/4$. From a field-theoretical point of view, this $AdS_4$ means that the symmetry-broken phase is described at low energies by a new IR CFT$_3$.

\section{Vortices as conformal defects} \label{sec:confdef}
We turn now to a discussion of the vortex. In a $2+1$ dimensional superfluid a vortex is a pointlike excitation around which the phase of the condensate winds. This means that the condensate $\langle \sO \rangle$ must vanish at the location of the vortex; this costs energy and will typically happen over a finite size, defining a core radius for the vortex. 

In our system the IR conformal invariance provides an extra ingredient. Note that there are two different CFTs, one in the UV and one in the IR. The UV conformal invariance is broken by the relevant double-trace coupling.
The UV theory is well-defined to arbitrarily high energy scales, and thus within this theory the vortex should be a normalizable and regular excitation.  In particular we expect it to have a finite energy and core radius set by the scale $\ka$ provided by the double-trace coupling. We will demonstrate this explicitly in later sections by constructing a gravitational description of the full vortex in this UV-complete theory. 

However in this section we will solve a simpler problem. Consider the infrared, i.e.  energies much smaller than $\ka$. From this point of view the vortex is an infinitely heavy and pointlike excitation, and thus corresponds to a {\it defect}, a non-normalizable modification of the IR CFT at a single point. At low energies the deformation should flow to a conformally invariant boundary condition at that point; thus we expect that the IR physics of these vortices can be understood from the theory of defect or boundary CFT \cite{Cardy:2004hm}. We first recall some basic concepts. 

Consider the IR CFT defined on $\mathbb{R}^{2,1}$ with a pointlike defect localized at the origin $\vec{x} = 0$ (and extending for all time). The CFT without the defect is invariant under the full conformal group $SO(3,2)$; this is broken down to $SO(2,1) \times SO(2)$ by the presence of the defect. $SO(2,1)$ is the symmetry group of a CFT$_1$ extending along the vortex worldline; thus one may say that there is a nontrivial CFT$_1$ living on the defect. 

The symmetry structure described above can be made more transparent if we perform a conformal rescaling to $AdS_2 \times S^1$:
\be
\dd s^2 = -\dd t^2 + \dd \rho^2 + \rho^2 \dd \varphi^2 = \rho^2\le(\frac{-\dd t^2 + \dd \rho^2}{\rho^2} + \dd \varphi^2\ri). \label{ads2confR}
\ee
The unbroken $SO(2,1) \times SO(2)$ now acts geometrically in an obvious fashion on $AdS_2 \times S^1$. The defect has been mapped to the boundary of $AdS_2$. The existence of the defect manifests itself in the need to specify boundary conditions at the $AdS_2$ boundary, and possibly around the non-contractible $S^1$. For a vortex it is clear that we should demand that the phase of the scalar condensate wind around this $S^1$. 

\subsection{Gravity solution}
We turn now to an explicit construction of the gravitational dual of this conformal defect. We first seek a suitable bulk coordinate system. Consider the line element of pure AdS$_4$ written in Fefferman-Graham coordinates:
\begin{equation}
\dd s^2 = \frac{\tilde{L}^2}{z^2}\left[-\dd t^2+\dd R^2+R^2\dd\varphi^2+\dd z^2\right].
\label{eq:pureAdS}
\end{equation} 
This $AdS_4$ is dual to the IR CFT, and so one should imagine it representing the IR portion of the geometry described in Section \ref{sec:setup}. There is no vortex here yet, but it is nevertheless helpful to imagine one sitting at the origin of field theory coordinates ($R = 0$) and hanging down into the bulk ($z$ arbitary). From this point of view one might think that the vortex solution will always depend on two coordinates $(R,z)$; as we now show, this is not true. 

Consider the following set of coordinates:
\begin{equation}
R= \rho\sin\theta \,,\quad\text{and}\quad z= \rho \cos\theta\,, \label{PoincareC}
\end{equation}
in terms of which the line element (\ref{eq:pureAdS}) reduces to
\begin{equation}
\dd s^2 = \frac{\tilde{L}^2}{\cos^2\theta}\left[\frac{-\dd t^2+\dd \rho^2}{\rho^2}+\dd \theta^2+\sin^2\theta\dd\varphi^2\right]\,. \label{eq:A2inA4}
\end{equation}
In these coordinates, pure AdS$_4$ is viewed as a warped fibration of AdS$_2$. Note that the conformal boundary of AdS$_4$ (located at $\th \to \frac{\pi}{2}$) is precisely $AdS_2 \times S^1$. Thus the dual CFT is defined on $AdS_2 \times S^1$, as in \eqref{ads2confR}, but this metric preserves the full $SO(3,2)$ isometry group of AdS$_4$. 

We argued above that a vortex breaks $SO(3,2)$ down to $SO(2,1) \times SO(2)$. The most general line element compatible with such symmetries is now
\begin{equation}
\dd s^2 = \frac{L^2}{\cos^2\theta}\left[F(\theta)\left(\frac{-\dd t^2+\dd \rho^2}{\rho^2}\right)+H(\theta)\dd \theta^2+G(\theta)\sin^2\theta\dd\varphi^2\right].
\label{eq:scaling}
\end{equation}
Both the functions in the metric $F(\th), G(\th), H(\th)$ and the matter sector $A_\varphi(\theta)$ and $\Phi(\theta)$ are functions of $\theta$ only.

As $\th \to 0$ we have the core of the vortex, where the $\varphi$ circle shrinks and the scalar and gauge field will vanish. As $\th \to \frac{\pi}{2}$ we approach the conformal boundary $AdS_2 \times S^1$; the metric functions approach those of $AdS_4$, and the matter fields satisfy the boundary conditions 
\be
\arg \Phi = n \varphi \qquad A_{\varphi}\le(\th \to \frac{\pi}{2}\ri) = \frac{n}{q} \,.\label{ads2bc}
\ee
 It is interesting that this solution depends only on a single coordinate $\th$ rather than $R$ and $z$ independently. We will see that the full solution (out to the UV $AdS_4$) does depend on two variables, but there is enhanced symmetry in the IR. This is a consequence of the conformal symmetry preserved by the vortex, essentially stating that moving away from the vortex is the same as moving deeper into the infrared.
 We will refer to this as the ``scaling solution".

There is an interesting property of the bulk metric \eqref{eq:scaling}; independent of the details of the metric functions, the existence of the $AdS_2$ endows the bulk solution with a Poincar\'e horizon at $\rho \to \infty$. There is an entropy associated with this horizon, which extends from $\th = 0$ to $\th = \pi/2$:
\be
S_H = \frac{ \pi L^2}{2 G_N} \int_0^{\th_\Lam} d\th \frac{\sin \th}{\cos^2\th} \sqrt{H(\th) G(\th)} \ . \label{Shor}
\ee
In this expression we have cut off the $\th$ integral at a value $\th_{\Lam} \sim \frac{\pi}{2}$. What is the precise interpretation of this entropy in the field theory? In the coordinates given by \eqref{eq:scaling}, this horizon intersects the conformal boundary $AdS_2 \times S^1$ at $\th = \frac{\pi}{2}$: thus in this conformal frame it can be viewed as a bulk minimal surface that hangs down from the boundary, and is computing a field-theoretical {\it entanglement entropy} via the Ryu-Takayanagi prescription \cite{Ryu:2006bv}. In fact any constant $\rho$ surface is a minimal surface, not just the surface as $\rho \to \infty$; furthermore they all have the same area, due to the $AdS_2$ isometry that shifts the value of $\rho$. The surface wraps the $S^1$: on the boundary this $S^1$ surrounds the defect, and thus we are computing the entanglement entropy of the defect with its surroundings. The analogous quantity in 2d CFT is called a {\it boundary entropy} as the defect there cuts the line into two, and is well-studied \cite{PhysRevLett.67.161}. We are not aware of much study in higher dimensions: however see the recent work \cite{Jensen:2013lxa}. 

There is a divergence in the expression \eqref{Shor} as we approach the boundary; this has nothing to do with the vortex and in the $AdS_2 \times S^1$ conformal frame may be interpreted as the usual UV divergence of the entanglement entropy. We may obtain a finite impurity entropy by subtracting the same entanglement entropy without the defect present, i.e. evaluating \eqref{Shor} on \eqref{eq:A2inA4}.
\be
S_{imp} = \lim_{\th_{\Lam} \to \frac{\pi}{2}}\le(S_H - \frac{ \pi \tilde L^2}{2G_N} \int_0^{\th_{\Lam}} d\th \frac{\sin \th}{\cos^2\th}\ri)\,. \label{impSdef}
\ee
$S_{imp}$ is a finite and universal number characterizing the defect.\footnote{Note that the entanglement entropies involved in this subtraction are defined in the $AdS_2 \times S^1$ conformal frame. One must impose the cutoff differently with the help of \eqref{PoincareC} to obtain entanglement entropies in the $\mathbb{R}^{2,1}$ conformal frame: in fact the value of $\th_{\Lam}$ then depends on $\rho$, introducing $\rho$-dependence in the value of the entanglement entropy.}

\subsection{\label{sec:scaling}Numerical construction}

We now discuss the explicit numerical construction of this geometry. It turns out to be convenient to work with a different angular coordinate:
\begin{equation}
\cos\theta = \tilde{y}\sqrt{2-\tilde{y}^2}\,,
\end{equation}
which brings the line element (\ref{eq:scaling}) to the following form
\begin{equation}
\dd s^2 = \frac{L^2}{\tilde{y}^2(2-\tilde{y}^2)}\left[F(\tilde{y})\left(\frac{-\dd t^2+\dd \rho^2}{\rho^2}\right)+\frac{4\,H(\tilde{y})\,\dd \tilde{y}^2}{2-\tilde{y}^2}+G(\tilde{y})(1-\tilde{y}^2)^2\dd\varphi^2\right],
\label{eq:ansatznear}
\end{equation}
where the vortex core is now located at $\tilde{y}=1$, and $\tilde{y}=0$ is the region infinitely far away from the vortex core.


The line element (\ref{eq:ansatznear}) still exhibits gauge freedom for arbitrary reparametrizations of $\tilde{y}$. In order to circumvent this problem (and its higher dimensional analog in the solution of partial differential equations in the next sections), we will use the DeTurck method, first introduced in \cite{Headrick:2009pv} and studied in great detail in \cite{Figueras:2011va}.

This is based on the so called Einstein-DeTurck equations, which can be obtained from the standard Einstein equations (\ref{eq:einsteinEOM}), by adding the following new term
\begin{equation}
G^{H}_{ab} \equiv G_{ab}-\nabla_{(a}\xi_{b)}=0,
\label{eq:einsteindeturck}
\end{equation}
where $\xi^a = g^{cd}[\Gamma^a_{cd}(g)-\bar{\Gamma}^a_{cd}(\bar{g})]$ and $\bar{\Gamma}(\bar{g})$ is the Levi-Civita connection associated with a reference metric $\bar{g}$. The reference metric is chosen to be such that it has the same asymptotics and horizon structure as $g$. This produces non-degenerate kinetic terms for all of the metric components and automatically fixes a gauge. Furthermore, the Einstein-DeTurck equation can be shown to be elliptic for static line elements \cite{Headrick:2009pv}\footnote{In fact, in \cite{Headrick:2009pv} it was shown that the Einstein DeTurck equations are elliptic under more general assumptions, but in this paper we only need the results regarding static line elements.}, such as the ones we consider in this manuscript.

It is easy to show that any solution to $G_{ab}=0$ with $\xi=0$ is a solution to $G^{H}_{ab}=0$. However, the converse is not necessarily true. In certain circumstances one can show that solutions with $\xi\neq 0$, coined Ricci solitons, cannot exist \cite{Figueras:2011va}. For the case at hand, we did not manage to prove such a theorem. Basically, the presence of the matter fields do not allow for a straightforward extension of proof given in \cite{Figueras:2011va}. However, since the equations we want to solve are elliptic, they can be solved as a boundary value problem for well-posed boundary conditions. The solutions to such equations can be shown to be locally unique. This means that a solution of the Einstein equations cannot be arbitrarily close to a DeTurck soliton, and that we should be able to distinguish between the two by monitoring $\xi_a \xi^a$. Note that for static line element, it can be easily shown that $\xi_a\xi^a>0$.

If we input the ansatz (\ref{eq:ansatznear}) into the Einstein-DeTurck equations (\ref{eq:einsteindeturck}) and matter field Eqs.~(\ref{eq:maxwellEOM}-\ref{eq:scalarEOM}), we find that the $\rho$ dependence cancels out and we are left with five second order nonlinear ODEs in $\tilde{y}$. This is not a surprise, since we have maintained $SO(2,1)$ symmetry. In our numerical code, we have decided to solve for the following set of variables $\{F(\tilde{y}),H(\tilde{y}),G(\tilde{y}),\widehat{A}_\varphi(\tilde{y}),\widehat{\Phi}(\tilde{y})\}$, where we defined
\begin{equation}
A_{\varphi}(\tilde{y})\equiv L (1-\tilde{y}^2)^2\widehat{A}_\varphi(\tilde{y})\quad\text{and}\quad \Phi(\tilde{y})\equiv(1-\tilde{y}^2)^n\,e^{i\,n\,\varphi}\widehat{\Phi}(\tilde{y})\,.
\end{equation}
Note that the factors of $(1-\tilde{y}^2)^2$ and $(1-\tilde{y}^2)^n$ in the definitions of $\widehat{A}_\varphi$ and $\widehat{\Phi}$, respectively, ensure that regularity at $\tilde{y}=1$ only requires pure Neumann boundary conditions on both $\widehat{A}_\varphi$ and $\widehat{\Phi}$. Furthermore, regularity of the line element (\ref{eq:ansatznear}) also demands $H(1)=G(1)$. The remaining boundary conditions at $\tilde{y}=1$ are of the pure Neumann type. These conditions can be obtained via an analysis similar to the one presented in detail later in Section \ref{sec:pdenum}. At $\tilde{y}=0$, we demand
\begin{equation}
\widehat{\Phi}(0)=1\,,\quad\widehat{A}_\varphi(0)=\frac{n}{q\,L}\quad\text{and}\quad F(0)=H(0)=G(0)=\frac{3}{4}\,.
\end{equation}
Note that the factors of $\frac{3}{4}$ here are due to the fact that the IR geometry without the vortex has an effective AdS radius of $\tilde{L}^2 \equiv \frac{3}{4} L^2$, as discussed in Section \ref{sec:setup}. Finally, for the DeTurck reference metric we chose $F(\tilde{y})=H(\tilde{y})=G(\tilde{y})=3/4$.

\begin{figure}[ht]
\centering
\includegraphics[width=0.95\textwidth]{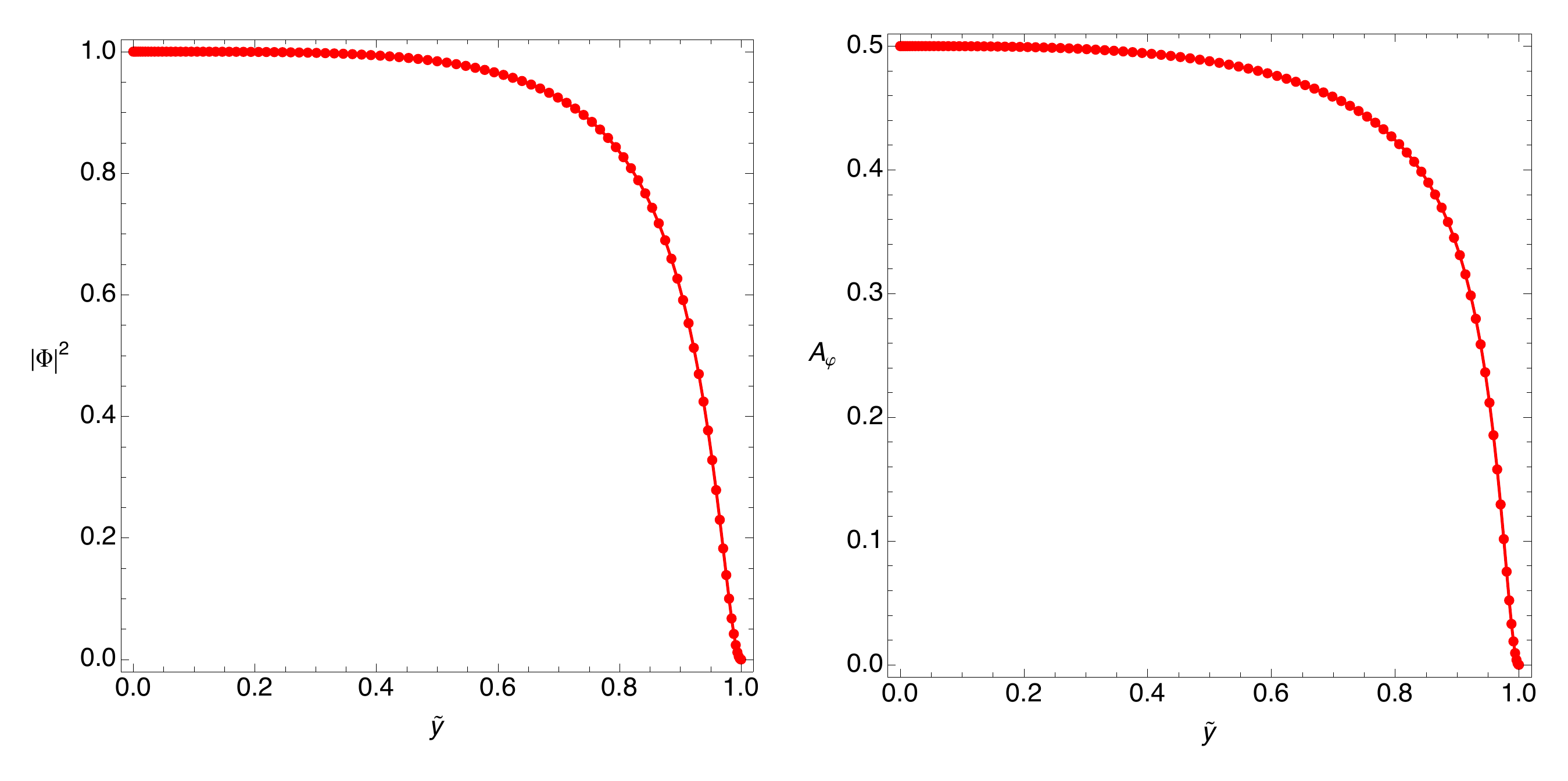}
\caption{Scalar field $|\Phi|$ ({\it left panel}) and gauge field $A_{\varphi}$ ({\it right panel}) as a function of $\tilde{y}$ for $q L = 2, n = 1$. The core of the vortex is at $\tilde{y} = 1$, where both functions must vanish by regularity. At $\tilde{y} \to 0$ we approach the homogeneous  ground state, and the scalar approaches the minimum of its potential. }
\label{fig:nearhorizonprof}
\end{figure}

We now present the results from this analysis for a vortex with $n = 1$. The resulting gauge field and scalar profiles are shown in Fig.~\ref{fig:nearhorizonprof} for $q L = 2$. We see that they interpolate smoothly from the core of the vortex at $\tilde{y} = 1$ to the IR CFT vacuum at $\tilde{y} = 0$. We stress that this is only an infrared limit of the vortex; in this approach the temperature is strictly zero, and we cannot include the irrelevant deformations that would take us eventually to the UV. 

\begin{figure}[ht]
\centering
\includegraphics[width=0.5\textwidth]{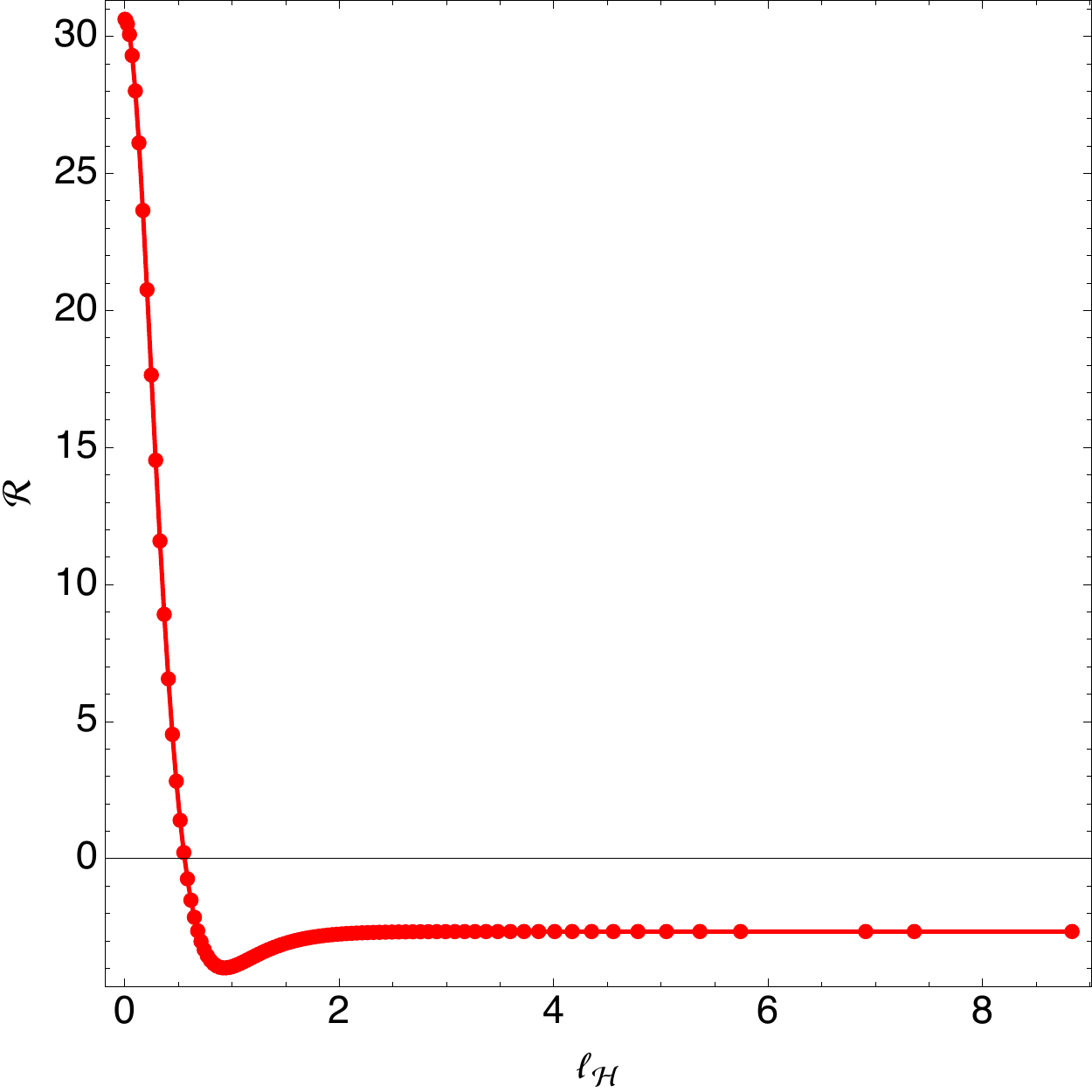}
\caption{The scalar curvature of the $T=0$ horizon as a function of proper distance from the vortex core. The large positive peak near the core denotes a ``bubble of Reissner-Nordstr\"{o}m horizon" sticking out of the usual Poincar\'e horizon. }
\label{fig:nearhorizoncurv}
\end{figure}

Since the solution is independent of $\rho$,  the geometry of the horizon at $\rho =\infty$ is the same as the geometry on any constant $\rho$ (and constant $t$) surface. Fig. \ref{fig:nearhorizoncurv}  shows the scalar curvature ${\cal R}$ of the horizon  as a function of proper distance from the vortex core. Note the large positive peak near the origin. This reflects a ``bubble of Reissner-Nordstr\"{o}m horizon" sticking out of the usual Poincar\'e horizon as we anticipated in the introduction. The fact that the curvature approaches a negative constant at large distance may seem puzzling, since one often thinks of the  Poincar\'e horizon in AdS as being flat. But that impression is incorrect, and results from extrapolating the Poincar\'e coordinates to the horizon where they are no longer valid. To see that a cross-section of the Poincar\'e horizon really has constant negative curvature, we can use \eqref{eq:A2inA4}: since the coordinate transformation \eqref{PoincareC} does not involve $t$, the horizon at $\rho = \infty$ is identical to the usual Poincar\'e horizon. The coordinates $(\theta,\varphi)$ are well defined there and parameterize a hyperbolic plane.
 
\begin{figure}[ht]
\centering
\includegraphics[width=0.5\textwidth]{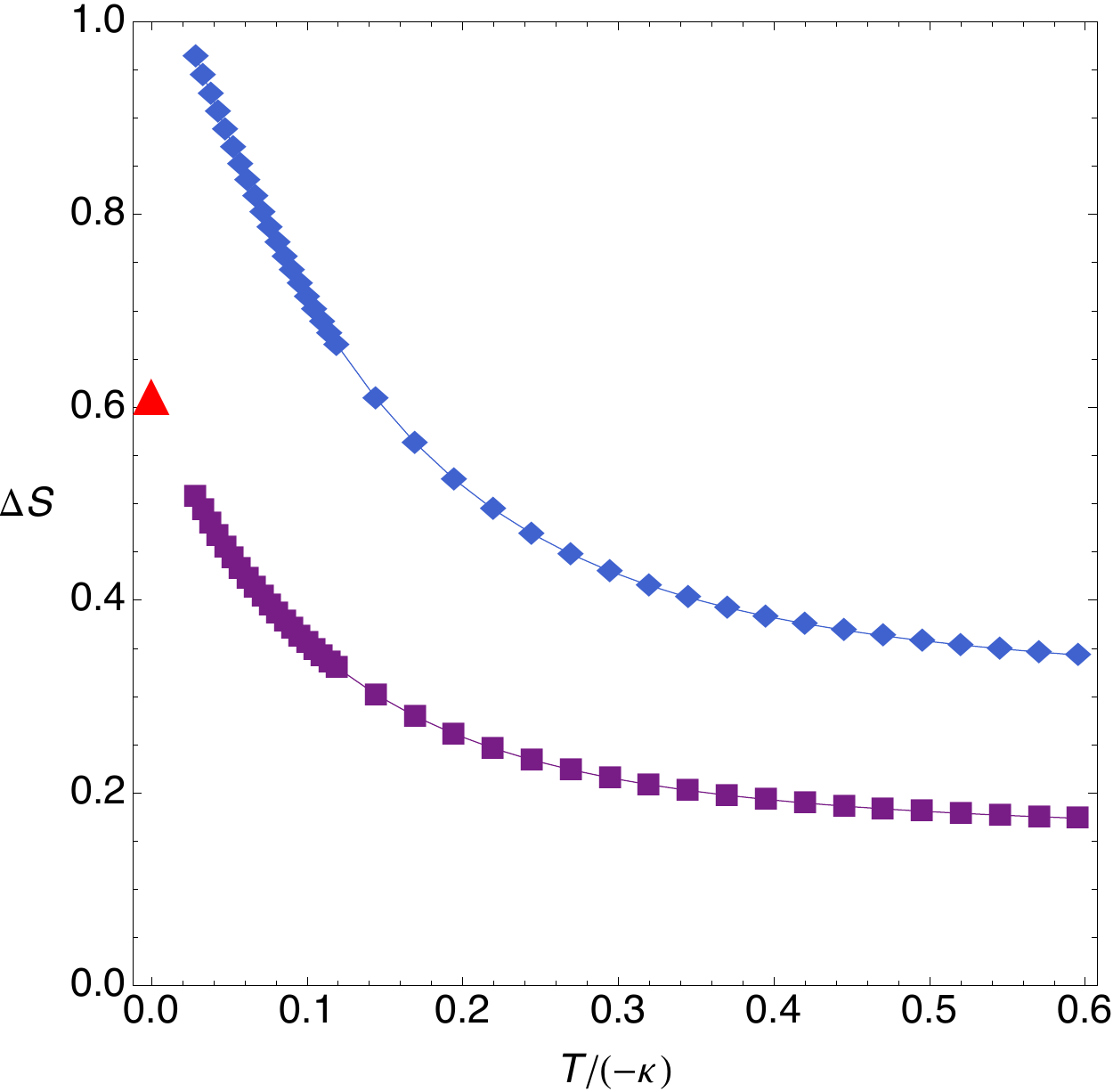}
\caption{Full entropy difference (defined later in \eqref{eq:difentropy}) as a function of $T/(-\kappa)$ for $q\,L =2$. Squares correspond to $n=1$ and diamonds to $n=2$. The red triangle represents the impurity entropy (defined in \eqref{impSdef}) extracted from the scaling solution \eqref{eq:ansatznear}.}
\label{fig:entropydiff}
\end{figure}

In the next sections we will solve the full partial differential equations describing the vortex at a nonzero $T$
 in the UV complete theory. When solving the PDEs it is technically difficult to work at precisely zero temperature. 
 Instead we will demonstrate that as we cool the vortex down various infrared observables computed from the full geometry appear to tend towards those arising from the scaling solution described in this section. 
 
We will discuss most of those results in Section \ref{sec:results} after describing their calculation, but to set the stage we present just one result in Fig.~\ref{fig:entropydiff}, where we compare the impurity entropy \eqref{impSdef} with the thermodynamic entropy difference $\Delta S$ of the full black hole with and without the vortex present, i.e.
 \be
 \Delta S(T) \equiv  S(T) - S_{0}(T) \ .
 \ee
We see that as we lower the temperature from finite but small temperatures to $T = 0$, the thermodynamic entropy $\Delta S$ appears to be in perfect agreement with the impurity entropy $S_{imp}$ that we find from the scaling solution. We see this as a very good indication that we have found the correct near horizon geometry. 

We turn now to a subtle point. In the field theory there are two natural definitions for a defect entropy: we can define a defect entropy at strictly zero temperature via the entanglement entropy of a symmetric region surrounding the defect, or consider instead the zero temperature limit of the thermodynamic entropy $\Delta S(T)$ defined above. In 2d CFT one can show on general grounds that these two definitions are equivalent \cite{Calabrese:2004eu,2009JPhA...42X4005C,2009JPhA...42X4009A}. This has also been directly verified in holographic calculations \cite{Azeyanagi:2007qj,Bak:2011ga}. In higher dimension this need not be the case, and indeed examples are known where these definitions disagree\footnote{An example is given by a probe string in $AdS_{d+1}$ with $d > 2$ \cite{Jensen:2013lxa}; we thank K. Jensen for drawing this to our attention.}. In our calculation we have taken the definition of the defect entropy
to be the regulated entanglement entropy
evaluated in the $AdS_2 \times S^1$ conformal frame \eqref{impSdef}, and 
 have shown that this matches very well with the $T \to 0$ limit of the thermodynamic entropy. While from the bulk point of view the subtraction involved in \eqref{impSdef} appears natural (in that we are subtracting the areas of two bulk horizons) the precise reason for this agreement from the field theory deserves further study.  

Another comparison one can make is between the impurity entropy and the entropy of an extreme Reissner-Nordstr\"{o}m solution with one unit of total flux. To define this latter quantity, one can start by compactifying the horizon into a finite volume torus. One finds that the entropy of the extremal solution is proportional to the magnetic flux. One can thus take the infinite volume limit and obtain a finite entropy. We have made this comparison and find that our impurity entropy is 
roughly double 
the Reissner-Nordstr\"{o}m entropy with the same total flux. 
Confining the flux into finite volume apparently increases its entropy.

We have not yet been able to construct the near horizon scaling solution for vortices with more than the minimum flux, i.e., $n>1$. So in section \ref{sec:results} we will only compare the finite temperature $n=1$
solutions to their $T=0$ limit.

\section{Full solution: boundary conditions and numerical methods} \label{sec:pdenum}

In this section we venture away from the infrared and describe the solution to the full problem of constructing a vortex in the UV complete theory. We demand that in the UV we approach the original $AdS_4$, with the scalar approaching the local maximum of its potential at $|\Phi| = 0$ and satisfying the double-trace boundary conditions \eqref{eq:double}. We will first explain the general ansatz used for determining both the metric and matter fields, and then discuss the appropriate  boundary conditions and numerical methods used to determine the solution. For convenience of notation we refer to the homogeneous superconducting black hole solution (to which our solutions asymptote in various limits) by the abbreviation HHH. 

\subsection{Metric and matter fields ansatz}
We want a configuration that, from the metric perspective, is symmetric under rotations about the origin of the vortex, so it is clear that we will have a rotation Killing vector $\partial_\varphi$. In addition, we are interested in static black hole solutions, which also means we will have a timelike Killing vector $\partial_t$. Finally, we expect the physics to depend both on the radial variable that measures the distance to the vortex core (we will call it $x$ or $R$) and on the holographic direction (which we denote as $y$ or $z$). So, we anticipate that our problem will be co-homogeneity two, and that cylindrical coordinates will be best adapted to study our problem.

The most general metric and matter ansatz compatible with the symmetries outlined above is:
\begin{subequations}
\begin{multline}
\dd s^2 = \frac{L^2}{y^2}\Bigg\{-Q_1\,y_+^2 (1-y^3)\dd t^2+\frac{Q_2\,\dd y^2}{1-y^3}+\\
\frac{y_+^2Q_4}{(1-x)^4}\left[\dd x+x\,y^2(1-x)^3\,Q_3\,\dd y\right]^2+\frac{y_+^2Q_5\,x^2}{(1-x)^2}\dd \varphi^2\Bigg\}\,,
\label{eq:ansatzmetric}
\end{multline}
\begin{equation}
\Phi =y\, e^{i\,n\,\varphi}\,x^n\,Q_6\qquad\text{and}\qquad A = L\,x^2 Q_7\,\dd \varphi\,,
\label{eq:ansatzgauge}
\end{equation}
\end{subequations}
where each of the $Q_i$'s is a function of $x$ and $y$ to be determined in what follows. For later numerical convenience, we have introduced several factors of $x$ and $y$ multiplying the functions $Q_i$.  Note that we write the phase of the complex scalar field as  $\tilde{\varphi}= n \,\varphi$  with $n$ being the winding number of the vortex along the Killing direction $\varphi$.  The eqs.~(\ref{eq:ansatzgauge}) are equivalent to the following gauge independent definitions
\begin{equation}
\Psi = y\, x^n\,Q_6\qquad\text{and}\qquad M_{\varphi} = L\,x^2 Q_7-\frac{n}{q}\,.
\end{equation}

In writing the solution in the above form, we have compactified both the radial distance from the vortex and the holographic direction. As a result, the coordinates $(x,y)$ take values in the unit square, with $y=1$ being the horizon location, $y=0$ the boundary at conformal  infinity, $x=0$ the core of the vortex and $x=1$ is asymptotic spatial infinity, i.e. infinitely far away from the vortex core. Regularity at the future and past horizons require all $Q_i$ to have a power series expansion in $(1-y)$, with $Q_1(x,1)=Q_2(x,1)$. 
It follows that the constant $y_+$ in (\ref{eq:ansatzmetric}) is proportional to the black hole Hawking temperature:
\begin{equation}
T = \frac{3\,y_+}{4\pi}\,.
\end{equation}
 As the dual theory is conformally invariant, the physics of each solution to the theory will then depend only on the dimensionless quantities $T/(-\kappa)$ (where $\kappa$ is given in (\ref{eq:double})), and on the vortex winding $n$. We will fix $\kappa = -1$ and use $y_+$ to probe different values of $T/(-\kappa)$.

The boundary conditions at $x=0$ are determined by  smoothness along the axis. The detailed conditions on the $Q_i$ are spelled out in Appendix \ref{app:bc}. The boundary conditions in the two asymptotic regions, $y=0$ and $x=1$, are a little subtle and will be discussed below.

The line element (\ref{eq:ansatzmetric}) still has gauge freedom associated with reparametrizations of $x$ and $y$. As before, we use the DeTurck method as introduced in \eqref{eq:einsteindeturck}, with the reference metric $\bar{g}$ given by the line element (\ref{eq:ansatzmetric}) with
\begin{subequations}
\begin{align}
&Q_1=Q_4=Q_5=1,\quad\text{and}\quad Q_3=0\,,
\\
&Q_2=1-\tilde{\alpha}\,y(1-y)\,,\label{eq:tildealpha}
\end{align}
\end{subequations}
where $\tilde{\alpha}$ is a constant that we will fix later.

\subsection{The holographic stress energy tensor and boundary conditions at the conformal boundary}

At the conformal boundary, located at $y=0$, we want our solution to approach AdS in Poincar\'e coordinates, i.e.
\begin{equation}
\dd s^2 = \frac{L^2}{z^2}\left(-\dd t^2+\dd R^2+R^2\dd \varphi^2+\dd z^2\right)\,.
\end{equation}
This implies Dirichlet boundary conditions for all metric functions, of the form
\begin{equation}
Q_1(x,0)=Q_2(x,0)=Q_4(x,0)=Q_5(x,0)=1\quad\text{and}\quad Q_3(x,0)=0\,,
\end{equation}
and the identification $R=x/(1-x)$, $y=y_+\,z$. Note that our reference metric $\bar{g}$ automatically satisfies these conditions.

The boundary conditions for the matter fields are better explained if we first introduce Fefferman-Graham coordinates $(z,\tilde{x})$ (FGC). Because we will determine all the $\{Q_i\}$ numerically, we can only hope to do this analytically close to the boundary. First we determine all the functions in an expansion in powers of $y$, by solving the equations off the boundary:
\begin{equation}
Q_i = \sum_{j=0}^{+\infty} Q^{(j)}_i(x)\,y^j\,,
\label{eq:Qs}
\end{equation}
where all the $Q^{(j)}_i(x)$ are determined as a function of $\{Q^{(3)}_4(x),Q^{(0)}_6(x),Q^{(1)}_6(x),Q^{(0)}_7(x),Q^{(1)}_7(x)\}$ and their derivatives along $x$.

A couple of comments are in order regarding this expansion. First, we have chosen the mass of our scalar field $\Phi$, namely $\mu^2L^2=-2$, to be such that near the conformal boundary
\begin{equation}
|\Phi| = |\tilde{\Phi}^{(1)}|\,y+|\tilde{\Phi}^{(2)}|\,y^2+\ldots\,\Rightarrow x^n\,Q_6 = |\tilde{\Phi}^{(1)}|+|\tilde{\Phi}^{(2)}|\,y+\ldots.
\label{eq:scalardecay}
\end{equation}
The boundary conditions presented in Section \ref{sec:setup} demand that $\tilde{\Phi}^{(2)}/\tilde{\Phi}^{(1)}$ is a constant, which translates into a Robin boundary condition for $Q_6$:
\begin{equation}
\left.\frac{\partial Q_6}{\partial y}\right|_{y=0}=\frac{\kappa_1}{y_+}Q_6(x,0)\,.
\label{eq:scalarBC}
\end{equation}
The precise relation between the double trace parameter $\kappa$ and $\kappa_1$ will be presented when discussing how to extract the holographic stress energy tensor. Note that the boundary condition for $Q_6$ is only of this simple form due to the extra factor of $y$ in the definition of $Q_6$ (see the first equation in Eq.~(\ref{eq:ansatzgauge})).

The second comment we want to make regarding the expansion (\ref{eq:Qs}) is that in general it will contain logarithms. However, that is not the case if one takes $\tilde{\alpha}=4\kappa_1/y_+$ in Eq.~(\ref{eq:tildealpha}), which we shall do from here henceforth. We have confirmed that this is the case, at least up to tenth order off the boundary.

Third, we need to discuss which boundary condition we impose on $A_\varphi$. This will depend on the physics we want to describe. The dual theory has a $2+1$ dimensional gauge coupling. If the gauge coupling is zero (so the $U(1)$ symmetry is not gauged)  we have a superfluid. As we do not want any external electromagnetic fields imposed,   $F_{\mu\nu}=0$ at the conformal boundary. This implies the following choice of boundary conditions

\begin{equation}
A_\varphi(x,0)=0\Rightarrow Q_7(x,0)=0\,.
\label{eq:BCsuperfluids}
\end{equation}
The other extreme is an infinite gauge coupling. This is a superconducting regime with zero current.
This means we should impose at the conformal boundary
\begin{equation}
\left.\frac{\partial M_\varphi}{\partial y}\right|_{y=0}=0\Rightarrow \left.\frac{\partial Q_7}{\partial y}\right|_{y=0}=0\,.
\label{eq:BCsupercondutors}
\end{equation}

From all the three comments above, we conclude that once the boundary conditions at the conformal boundary for $\Phi$ and $A_\varphi$ are suitably imposed, the $Q^{(j)}_i(x)$ are determined as functions of $\{Q^{(3)}_4(x),Q^{(0)}_6(x),\eta Q^{(0)}_7(x)+(1-\eta)Q^{(1)}_7(x)\}$ and their derivatives, where $\eta = 1$ for superconductors, and $\eta = 0$ for superfluids, i.e. a total of three functions in each phase. At this stage we would like to understand what is the physical meaning of such functions. This is best understood if we first change to FGC. We do this in an expansion off the boundary, by setting:
\begin{subequations}
\label{eqs:FGcoordiantes}
\begin{equation}
\left\{
\begin{array}{l}
\displaystyle y=y_+\,z+\sum_{j=2}^{+\infty} a_j(\tilde{x})z^j\,,
\\
\\
\displaystyle x=\tilde{x}+\sum_{j=1}^{+\infty} b_j(\tilde{x})z^j\,,
\end{array}
\right.
\end{equation}
and demanding that in the $(z,\tilde{x})$ coordinates $g_{zz}=L^2/z^2$ and $g_{z\tilde{x}}=0$. Note that at each order in $z$, we have two conditions to be solved for the two functions $\{a_j(\tilde{x}),b_j(\tilde{x})\}$. For completeness we provide here the first few terms in the above expansion:
\begin{align}
&a_2(\tilde{x})=\frac{5 \kappa _1 y_+}{4}\,,\\
&a_3(\tilde{x})=\frac{ \kappa_1 y_+}{64} \left(265\,\kappa_1-64\,y_+\right)\,,\\
&a_4(\tilde{x})=-\frac{ y_+ }{768}\left[336\,\kappa_1\,y_+^2\,\tilde{x}^{2 n}\,Q^{(0)}_6(\tilde{x})^2-14625\,\kappa_1^3+6720\,\kappa_1^2\,y_++128 y_+^3\right]\,,\\
&b_1(\tilde{x})=b_2(\tilde{x})=b_3(\tilde{x})=0\,,\\
&b_4(\tilde{x})=-\frac{y_+^2}{8}(1-\tilde{x})^4 \tilde{x}^{2\,n-1} Q^{(0)}_6(\tilde{x}) \left[\tilde{x}\,Q^{(0)}_6{}^\prime(\tilde{x})+n\,Q^{(0)}_6(\tilde{x})\right]\,.
\end{align}
\end{subequations}

We are now ready to explain the physical meaning of $Q^{(0)}_6(\tilde{x})$ and the relation between $\kappa_1$ and the usual double trace parameter $\kappa$. $\kappa$ is usually defined with respect to the FGC in the following way $-$ see \eqref{eq:asympscalar} and \eqref{eq:double} $-$ 
\begin{equation}
\Phi = \Phi^{(1)} z+\kappa\,\Phi^{(1)} z^2+\ldots\,,
\label{eq:kappajorge}
\end{equation}
with $\Phi^{(1)}$ being identified as the expectation value of the operator dual to $\Phi$, i.e.
\begin{equation}
\Phi^{(1)} =y_+\,e^{i\,n\,\varphi}\,Q^{(0)}_6(\tilde{x})\tilde{x}^n \equiv \langle \mathcal{O}\rangle\Rightarrow \left|\langle \mathcal{O}\rangle\right| = y_+\,\left|Q^{(0)}_6(\tilde{x})\right|\tilde{x}^n\,.
\end{equation}
Using both Eq.~(\ref{eq:scalardecay}) and Eq.~(\ref{eq:scalarBC}), together with the relation between $y$ and $z$ described in Eqs.~(\ref{eqs:FGcoordiantes}), we find
\begin{equation}
\kappa_1 = \frac{4\,\kappa}{9}\,.
\end{equation}

A similar expansion holds close to the conformal boundary for the gauge field $A_\varphi$, namely
\begin{equation}\label{eq:bdryA}
A_\varphi = L\,A^{(0)}_\varphi+L\,A^{(1)}_\varphi\,z+\ldots\,,
\end{equation}
where, acording to the usual AdS/CFT dictionary, $A^{(0)}_\varphi$ is the boundary Maxwell field, 
$A^{\mathrm{FT}}_\varphi$, and $A^{(1)}_\varphi$ the current flowing along $\varphi$, $J_\varphi$. This then implies the identification
\begin{equation}\label{eq:bdryAJ}
A^{\mathrm{FT}}_\varphi = \tilde{x}^2\,Q^{(0)}_7,\quad \text{and}\quad J_\varphi=y_+\,\tilde{x}^2\,Q^{(1)}_7\,.
\end{equation}

We are thus left to find an interpretation for $Q^{(3)}_4(\tilde{x})$. Not surprisingly, this will be related to the holographic stress energy tensor, whose extraction from the numerical data we detail below. Here we have decided to use the approach described in \cite{Balasubramanian:1999re}, and reconstruct the holographic stress energy tensor as
\begin{equation}
T_{\mu\nu} = \frac{1}{8\pi G_N\,L^2}\lim_{z\to 0}\left(\frac{L}{z}\right)\left(K_{\mu\nu}-\gamma_{\mu\nu}K-\frac{2}{L}\gamma_{\mu\nu}+L\,G^{(3)}_{\mu\nu}-\frac{\Phi^2}{L}\gamma_{\mu\nu}\right)\,,
\label{eq:stressenergy}
\end{equation}
where Greek indices run over boundary coordinates, $K_{\mu\nu}$ is the extrinsic curvature associated with an inward unit normal vector to the boundary (located at $z=0$), $K\equiv \gamma^{\mu\nu}K_{\mu\nu}$, $\gamma_{\mu\nu}$ is the induced metric on the constant $z$ surface, and $G^{(3)}_{\mu\nu}$ is the Einstein tensor of $\gamma_{\mu\nu}$. Since we are interested in field theories living on Minkowski space, the fourth term in Eq.~(\ref{eq:stressenergy}) vanishes as $z\to0$. The last term, on the other hand, gives the necessary contribution to cancel the divergences arising due to the presence of the scalar field. Note also that we used FGC to define the holographic stress energy tensor.

This stress energy tensor can be shown to obey  the following relations
\begin{equation}
h^{\mu\nu}T_{\mu\nu}=\frac{\kappa}{4\pi G_N}|\langle\mathcal{O}\rangle|^2\quad\text{and}\quad \tilde{\nabla}^\mu T_{\mu\nu}=\frac{\kappa}{8\pi G_N}\tilde{\nabla}_\nu |\langle\mathcal{O}\rangle|^2+\frac{1}{8\pi G_N}F^{\mathrm{FT}}_{\nu\rho}J^\rho\,,
\label{eq:conservation}
\end{equation}
where $h_{\mu\nu}$ is the metric on the conformal boundary, and $\tilde{\nabla}$ its associated Levi-Civita connection. For the solutions we are considering, the last term does not contribute, because superfluid boundary conditions require $F^{\mathrm{FT}}=0$, whereas our superconducting boundary conditions require $J=0$. The stress energy tensor would apriori depend on three unknown functions, $\{Q^{(3)}_1(\tilde{x}),Q^{(3)}_4(\tilde{x}),Q^{(3)}_5(\tilde{x})\}$, however, by using both conditions above, we can solve algebraically for $Q^{(3)}_1(\tilde{x})$ and $Q^{(3)}_5(\tilde{x})$, in terms of $Q^{(3)}_4(\tilde{x})$ and its derivatives along $\tilde{x}$.

A useful test of the numerics is given by the first law, expressed in the canonical ensemble variables:
\begin{equation}
\dd F = -S\,\dd T\,,
\label{eq:firstlaw}
\end{equation}
where $F=E-TS$ is the Helmoltz free energy. In order to use the above differential equations, we first need to define energy for these systems. This seems rather hopeless, because the holographic stress energy tensor is not covariantly conserved, see the second equation in (\ref{eq:conservation}). However, as we mentioned before, the last term in Eq.~(\ref{eq:conservation}) does not contribute on our solutions, and because the first term on the right hand side of the conservation equation is a total derivative, we can readily reabsorb it into an effective stress energy tensor, $\tilde{T}_{\mu\nu}$, that is conserved on our solutions:
\begin{equation}
\tilde{T}_{\mu\nu}=T_{\mu\nu}-\frac{\kappa\,h_{\mu\nu}}{8\pi G_N}|\langle\mathcal{O}\rangle|^2\,.
\end{equation}
We now define the energy in the usual way:
\begin{equation}
E = -\int_{\Sigma_t} \dd^2 x\sqrt{\eta}\,\tilde{T}_{\mu\nu}(\partial_t)^\mu t^\nu\,,
\label{eq:energystress}
\end{equation}
where $\eta_{\mu\nu}$ is the induced metric on the constant $t$ surface with unit normal $t^{\nu}$. Next we discuss the boundary conditions far from the vortex core.

\subsection{\label{sec:x1}Boundary conditions infinitely far away from vortex core - $x=1$:}
Because the flux is conserved, the boundary conditions at $x=1$ are very distinct for the superconductor and superfluid phases. Let us see why. The flux through a surface $\Sigma$ at constant $t$ and $y$ 
is given by
\begin{equation}
\tilde{\Phi} = \int_{\Sigma}  F\,.
\label{eq:flux1}
\end{equation}
For an isolated vortex, this flux is quantized and given by
\begin{equation}
\tilde{\Phi} =2\,\pi\, \frac{n}{q}\,.
\label{eq:flux2}
\end{equation}

Let us first start with the superconducting phase. This phase is characterized by a ``no current'' boundary conditions, see Eq.~\eqref{eq:BCsupercondutors}. This means that the field lines of $A_{\varphi}$ are allowed to penetrate the boundary, and we expect the magnetic field to fall off quickly far away from the vortex core. In particular, there is no tension between Eq.~(\ref{eq:flux1}) and Eq.~(\ref{eq:flux2}). In this case, the solution approaches the HHH solution \cite{Faulkner:2010gj} as we approach $x=1$.\footnote{Note that in the solution \cite{Faulkner:2010gj}  both the scalar field phase and the Maxwell field vanish while our solution has a complex scalar and $A_\varphi\neq 0$. However, the gauge transformation $\tilde{\varphi}\to \tilde{\varphi} +q \chi\,,\: A_\varphi \to A_{\varphi} +\nabla_\varphi \chi$, with gauge parameter $\chi=n\varphi/q$ rewrites the solution  of \cite{Faulkner:2010gj}  in the form \eqref{eq:HHHbc}.}
 The boundary conditions are simply
\begin{align}\label{eq:HHHbc}
&Q_1(1,y)=\tilde{Q}_1(y)\,\quad Q_2(1,y)=\tilde{Q}_2(y),\quad Q_3(1,y)=0\,,\nonumber
\\
\\
&Q_4(1,y)=Q_5(1,y)=\tilde{Q}_3(y)\,,\quad Q_6(1,y) = \tilde{Q}_4(y),\quad\text{and}\quad Q_7(1,y) = \frac{n}{q\,L}\,,\nonumber
\end{align}
where $\tilde{Q}_i(y)$ corresponds to the HHH solution expressed in DeTurck like coordinates.

Things are different if we instead consider the superfluid phase. Here, because $A_\varphi = 0$ at the boundary, there seems to be a tension between Eq.~(\ref{eq:flux1}) and Eq.~(\ref{eq:flux2}). This conundrum is solved in a simple but very dramatic way, namely, the field lines of $A_{\varphi}$ spread as we reach the boundary, and accumulate at $(x,y)\sim(1,0)$. This accumulation ends up destroying the asymptotics of the would be HHH black hole and creating a new solution that is similar to the usual holographic superconductor, except that $A_\varphi$ now depends on $y$, being $0$ at $y=0$, and approaching a smooth nonzero value at $y=1$. Close to $y=1$ we expect this new solution to be similar to the usual HHH black hole. Unlike the HHH solution, this black hole is not expected to exist as a solution of the Abelian-Higgs model in AdS \emph{per se}. Instead, it only makes sense as an asymptotic solution valid close $x=1$. One easy way of noting this is that this solution is not regular everywhere in our manifold, being singular if continued all the way to $x=0$, i.e. it violates the boundary conditions described in Appendix \ref{app:bc}. To sum up, the boundary conditions close to $x=1$ take the following form
\begin{align}
&Q_1(1,y)=\tilde{Q}_1(y)\,\quad Q_2(1,y)=\tilde{Q}_2(y),\quad Q_3(1,y)=0\,,\nonumber
\\
\\
&Q_4(1,y)=Q_5(1,y)=\tilde{Q}_3(y)\,,\quad Q_6(1,y) = \tilde{Q}_4(y),\quad\text{and}\quad Q_7(1,y) = \tilde{Q}_5(y)\,,\nonumber
\end{align}
where the $\tilde{Q}_i$ are now determined by solving five coupled ODEs that are obtained as limiting equations of our general PDE system as one approaches $x=1$. Finally, the factor of $(1-x)^3$ in the cross term of the line element (\ref{eq:ansatzmetric}) was chosen such that $Q_3$ vanishes \emph{linearly} at $x=1$, i.e. $Q_3\propto (1-x)$. The fact that $Q_3$ is linear in $(1-x)$, rather than some other higher power of $(1-x)$, is important to achieve the desired numerical accuracy.


\subsection{Numerical method and convergence}
Before proceeding to the discussion of the results we will first give some details on the numerical methods we employed. We have used a pseudospectral collocation procedure to descretize our PDE system. For both the $x$ and $y$ directions we used a collocation grid on Gauss-Chebyshev-Lobbato points. We solved the resulting system of nonlinear algebraic equations using a standard Newton-Raphson method.

\begin{figure}[ht]
\centering
\includegraphics[width=0.5\textwidth]{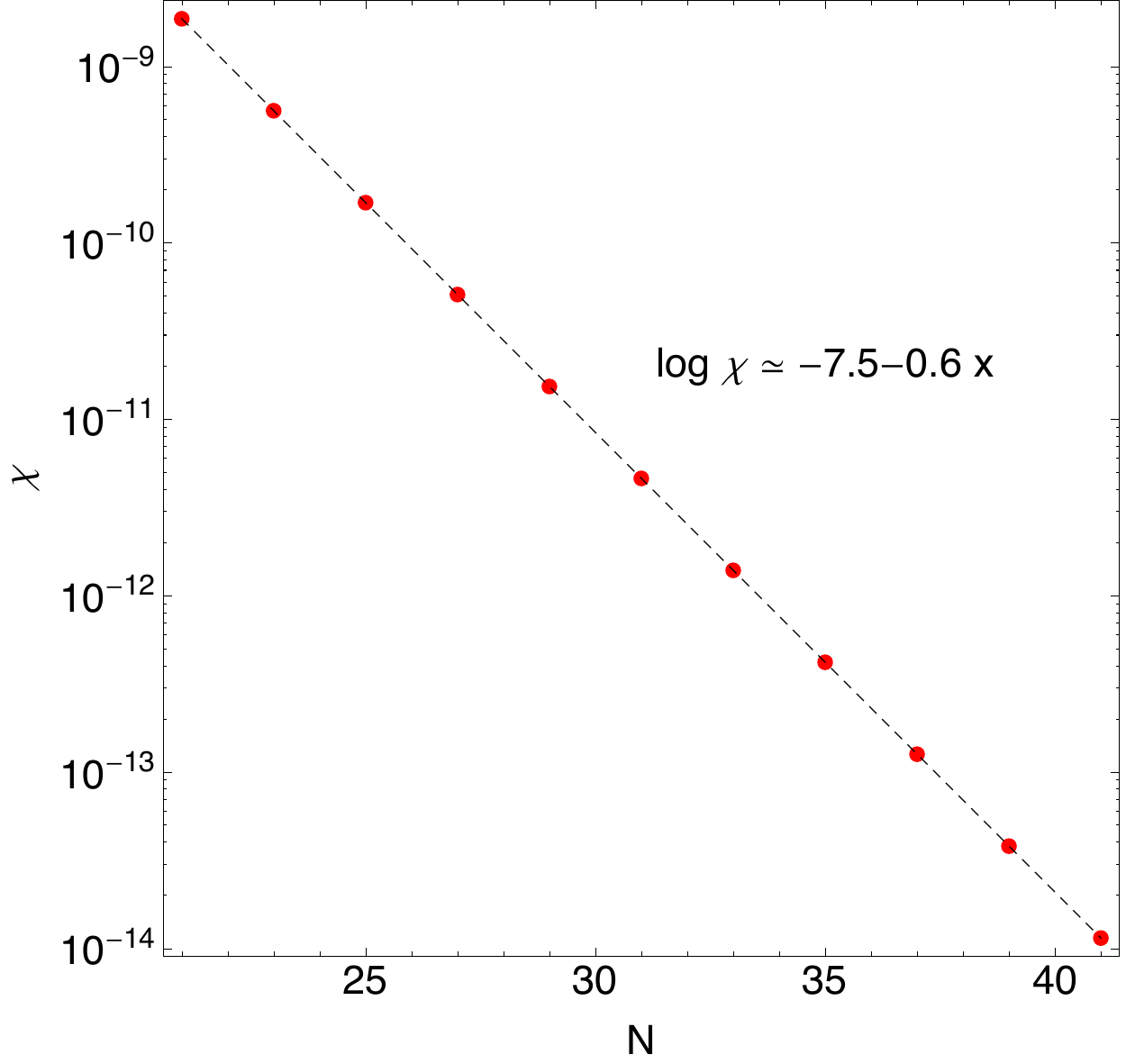}
\caption{$\chi_N$ as a function of the number of grid points $N$. The vertical scale is logarithmic, and the data is well fit by an exponential decay: $\log\chi_N = -7.5-0.60\,N$. In this particular simulation we have used $y_+ = 1/2$, $\kappa = -1$, $q=2$ and $n=1$.}
\label{fig:convergence}
\end{figure}

We have developed several tests of the convergence of our numerical method. First, we monitored the maximum of the norm of the DeTurck vector, as a function of the number of collocation points $N$, i.e. $\chi_N = \max_{(x,y)\in(0,1)^2} \xi_a \xi^a$. Note that we expect the norm to be zero on solutions of the Abelian Higgs model in AdS. Furthermore, since we are using a pseudospectral method in a Chebyshev grid, we expect the convergence of our numerical method to be exponential. We have confirmed that this is the case, as can be seen in Fig.~\ref{fig:convergence}.
We have also tested convergence by looking at how quantities such as the energy and entropy vary when we vary $N$, and we always find exponential convergence both for the superconducting and superfluid phases. A couple of remarks are in order about convergence across the parameter range we have probe. First, we note that as we lower the temperature, i.e. small values of $y_+$, we need to increase the number of points in order to keep the resolution. For instance, in order for $\chi_N$ to drop bellow $10^{-10}$ in the superconducting phase for $y_+ = 1/10$ we had to use $61$ grid points in both the $x$ and $y$ grids. Second, all computations in this paper were done using quadruple precision.

Finally, we have also tested the first law, Eq.~(\ref{eq:firstlaw}). We find perfect agreement, i.e. deviations under the percent level, when $y_+\gtrsim0.2$. However, when $y_+\sim 0.1$ we find deviations from this expression up to $5\%$. As we have mentioned before, this is not surprising since low temperature solutions are more difficult to determine accurately. Presumably, this difficulty is related with the fact that a throat is developing as we lower the temperature and that a more dense grid is required in order to resolve it. We note, however, that as we increase the number of points, Eq.~(\ref{eq:firstlaw}) is more and more accurate, with the expected exponential convergence for \emph{all} values of $y_+$ we simulated, namely $y_+\in(0.1,2.6)$. This roughly corresponds to the interval $T/(-\kappa)\in (0.023,0.621)$.

\section{Full solution: results} \label{sec:results}
In this section we present the results for the holographic duals of isolated vortices. 
We discuss superconducting and superfluid vortices separately: while these are similar in some ways (e.g. the physics of the horizon is very similar for both), certain aspects of the physics are very different.

\subsection{Superconducting vortices}

\subsubsection{Horizon properties} \label{sec:sc_hor}

As we discussed in section III, a novel feature of holographic vortices is that they carry magnetic flux out of the black hole horizon, distorting it. At low temperature, the horizon  approaches the Poincar\'e horizon of the IR $AdS_4$ away from the vortex, but at the core of the vortex, the scalar field vanishes and there is a single unit of nonzero magnetic flux. As a result, there is a piece of local Reissner-Nordstr\"{o}m AdS horizon inside the vortex, carrying a finite amount of entropy. 

To illustrate this ``horizon bubble" sticking out, we start with a diagram showing an isometric  embedding of the horizon into 3D hyperbolic space. 
Using the line element (\ref{eq:ansatzmetric}), one finds that the induced line element on the vortex horizon is given by
\begin{equation}
\dd s^2 = L^2\left[\frac{y_+^2Q_4(x,1)\,\dd x^2}{(1-x)^4}+\frac{y_+^2\,Q_5(x,1)\,x^2}{(1-x)^2}\dd \varphi^2\right]\,.
\label{eq:inducedhorizon}
\end{equation}
To construct an embedding diagram, one starts with the line element of hyperbolic space:
\begin{equation}
\dd s^2_{\mathbb{H}}=\frac{L^2}{\hat{z}^2}\left[\dd \hat{R}^2+\hat{R}^2\dd\varphi^2+\dd\hat{z}^2\right]\,.
\end{equation}
One then searches for an embedding of the form $(\hat{R}(x),\hat{z}(x))$, which gives the following metric
\begin{equation}
\dd s^2_{\mathbb{H}}=\frac{L^2}{\hat{z}(x)^2}\left\{\left[\hat{R}'(x)^2+\hat{z}'(x)^2\right]\dd x^2+\hat{R}(x)^2\dd\varphi^2\right\}\,.
\end{equation}
By equating the above line element, with Eq.~(\ref{eq:inducedhorizon}), one finds that
\begin{equation}
\frac{\hat{R}'(x)^2+\hat{z}'(x)^2}{\hat{z}(x)^2}=\frac{y_+^2Q_4(x,1)}{(1-x)^4}\quad \text{and}\quad \frac{\hat{R}(x)^2}{\hat{z}(x)^2}=\frac{y_+^2\,Q_5(x,1)\,x^2}{(1-x)^2}\,.
\end{equation}
These are first order nonlinear equations in $\hat{R}(x)$ and $\hat{z}(x)$ that can be readily solved using pseudospectral collocation methods. We fix the integration constants by demanding $\hat{z}(1)=1/y_+$. The curve traced by $(\hat{R}(x),\hat{z}(x))$, as we vary $x$ in the interval $(0,1)$, is the embedding diagram.

\begin{figure}[ht]
\centering
\includegraphics[width=0.5\textwidth]{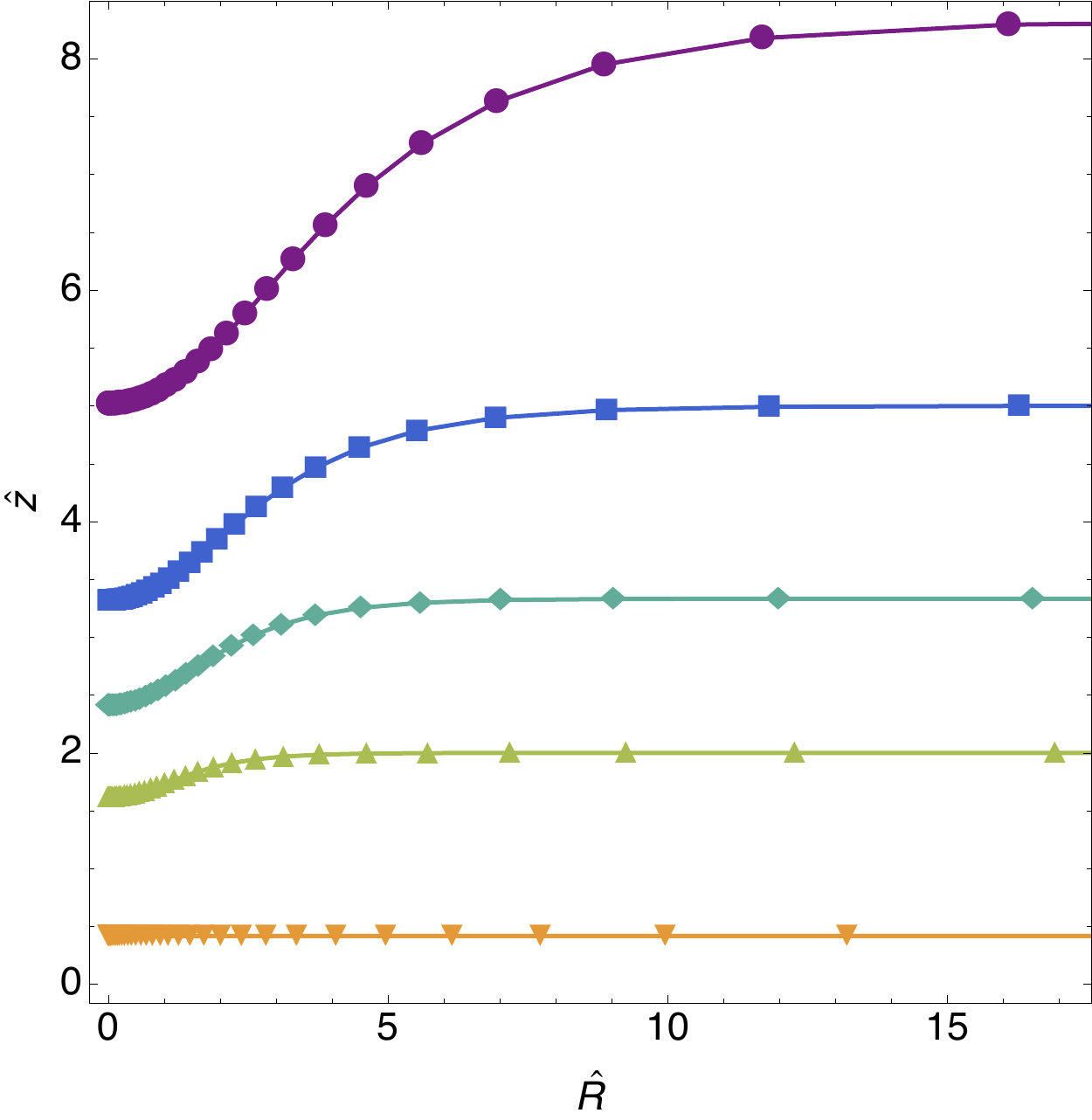}
\caption{Embedding diagram, plotted for several temperatures, for superconductor boundary conditions with $q\, L=2$ and $n=1$. Disks, squares, diamonds, triangles and inverted triangles have $T/(-\kappa) = 0.029,\, 0.048,\,0.072,\,0.119,\,0.571$, respectively.}
\label{fig:embeddingsc}
\end{figure}

The results for several different temperatures are shown in 
 Fig.~\ref{fig:embeddingsc}, where the temperature decreases from bottom to top. We see that as the temperature is decreased, the backreaction on the metric is such that a bulge is created on the horizon -- recall that smaller $z$ is closer to the conformal boundary. Asymptotically, i.e.
  as $\hat R(x)\to+\infty$, the horizon becomes flat. We have plotted this diagram for several values of $n$, and they all look qualitatively similar. Since the horizon is bulging out, one might  worry about a possible Gregory-Laflamme type instability on the horizon \cite{Gregory:1993vy}. We have not yet performed an exhaustive study of stability of this background, but our preliminary results indicate stability for the $n=1$ mode (see later discussion). 

\begin{figure}[ht]
\centering
\includegraphics[width=\textwidth]{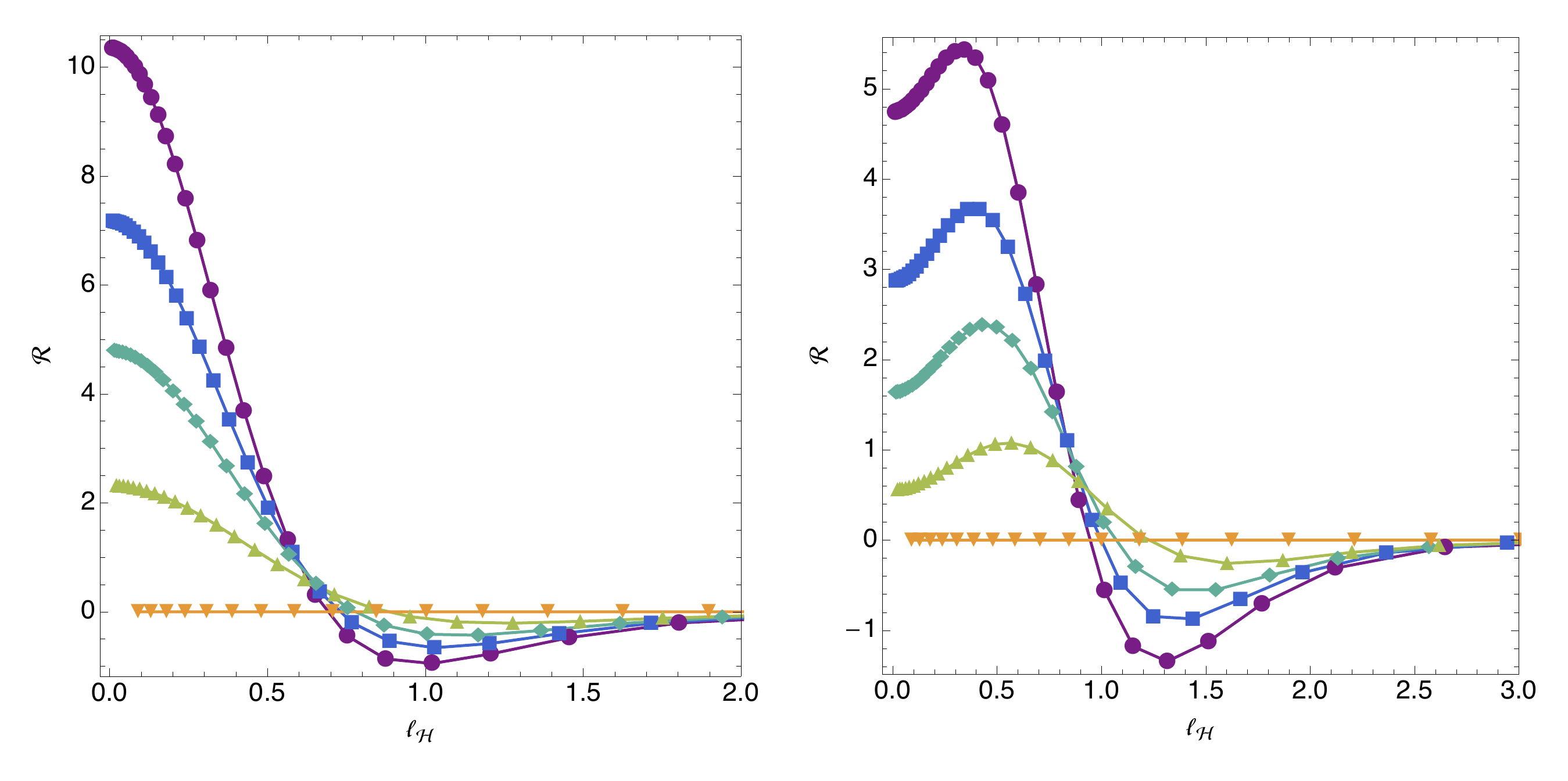}
\caption{Ricci scalar of the induced metric on the horizon, ${\cal R}$, for superconductor boundary conditions with $q\, L=2$, as a function of the proper distance to the vortex origin $\ell_\mathcal{H}$. The {\it left panel} has $n=1$, and the {\it right panel} $n=2$. In both panels, disks, squares, diamonds, triangles and inverted triangles have $T/(-\kappa) = 0.029,\, 0.048,\,0.072,\,0.119,\,0.571$, respectively.}
\label{fig:scalarHn1n2sc}
\end{figure}

\begin{figure}[h]
\centering
\includegraphics[width=0.5\textwidth]{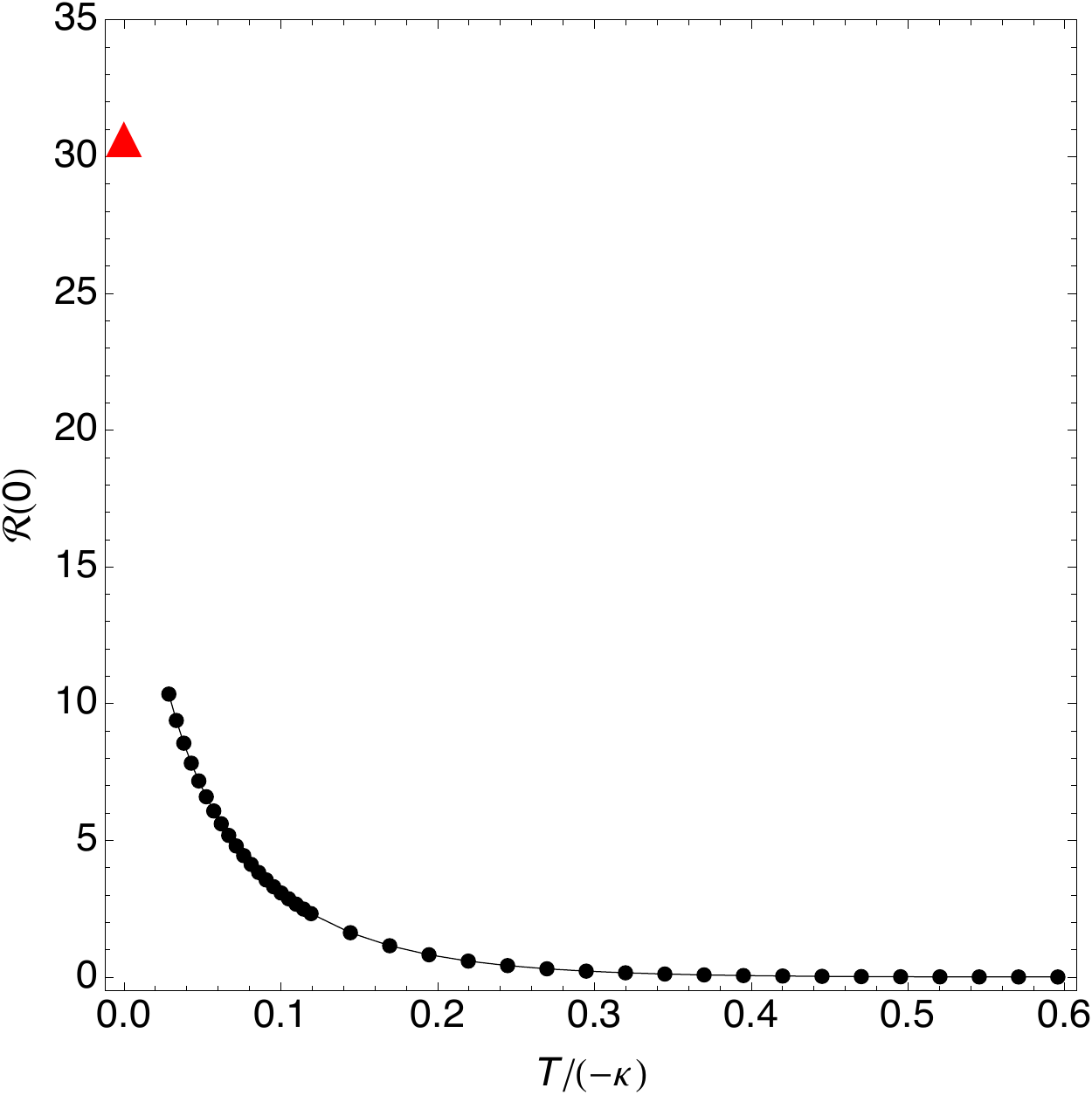}
\caption{Ricci scalar of the induced horizon geometry, evaluated at the origin, for superconductor boundary conditions with $q\, L=2$ and $n=1$, as a function of the temperature. The red triangle indicates the $T = 0$ result from the scaling solution constructed in section \ref{sec:confdef}.}
\label{fig:scalarHsc}
\end{figure}

We have also computed the Ricci scalar, ${\cal R}$, along the horizon, i.e. the Ricci scalar of the line element (\ref{eq:inducedhorizon}), as a function of proper distance, $\ell_H$, from the rotation axis. The results for vortices with 
 $n=1$ or $n=2$ are shown in Fig.~\ref{fig:scalarHn1n2sc}.\footnote{In this, and all subsequent bulk plots, we have set $L = 1$ so everything is measured in units of the AdS radius.} For $n=1$ we find that the maximum always sits at the origin, whereas for $n=2$ it is obtained around $\ell_H\sim1/2$. This shift is simply a consequence of the fact that the energy density caused by the complex scalar has two main contributions: $g^{xx}|\nabla_x\Phi|^2+g^{\varphi\varphi}|\nabla_\varphi\Phi|^2$. 
 Finally, in all cases we see that the Ricci scalar approaches $0$ as $\ell_H \to+\infty$, since we recover the HHH black hole, which has an horizon that preserves translational invariance.

In Fig.~\ref{fig:scalarHsc} we plot the Ricci scalar for $n=1$ evaluated at the origin. Note that it increases monotonically as the temperature is decreased; we also indicate the precise value at $T = 0$, obtained from Fig. \ref{fig:nearhorizoncurv}.
 While this value does fit the trend, we see that we are still some distance away from zero temperature. Note the utility of the scaling solution; without it, the steep upwards slope might have made us nervous that our $T > 0$ solutions would have a singular $T = 0$ limit. 

\begin{figure}[h]
\centering
\includegraphics[width=\textwidth]{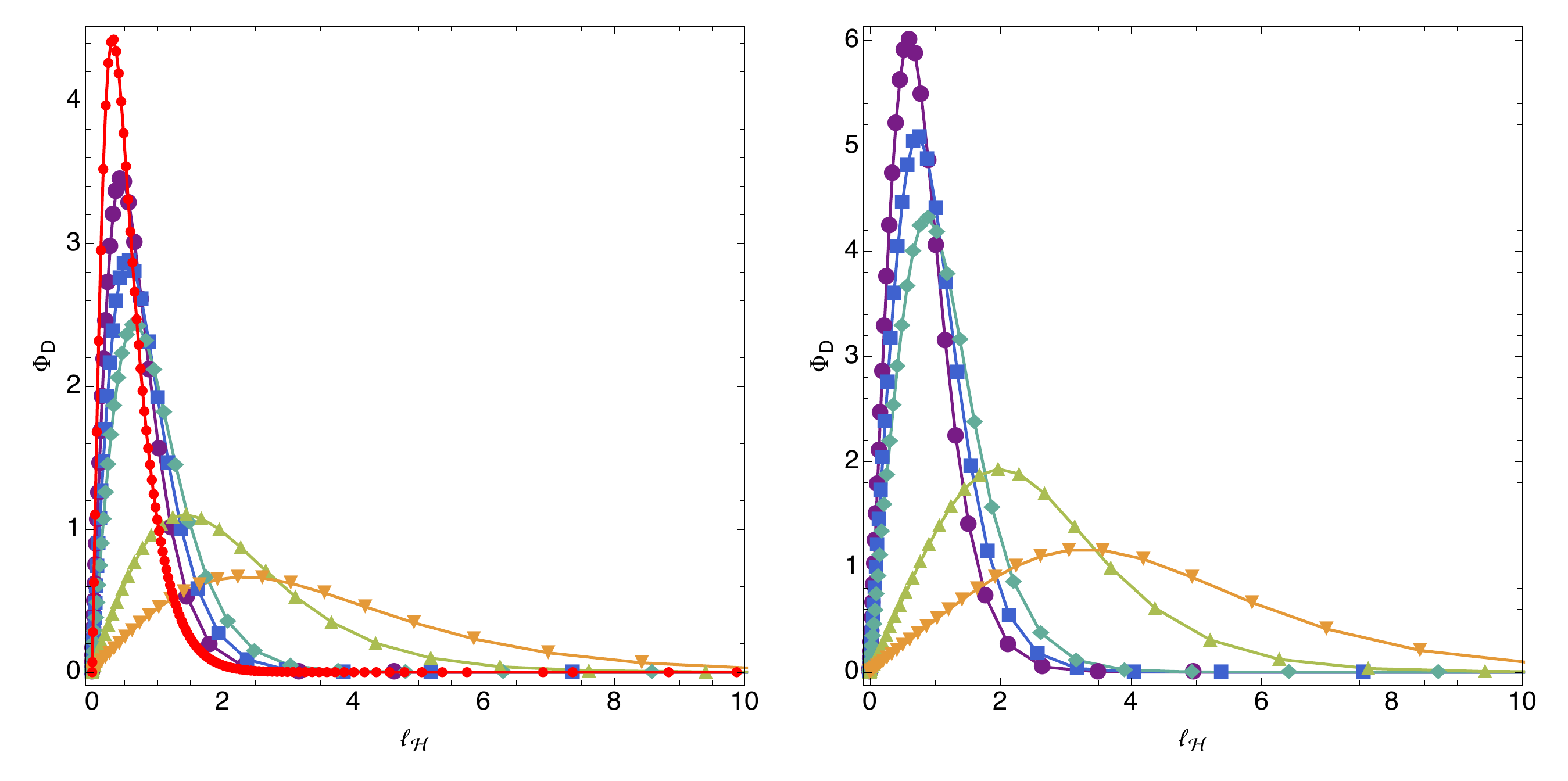}
\caption{Magnetic flux density \eqref{eq:fluxdensity} evaluated at the horizon, as a function of the proper distance along the horizon, plotted for several temperatures.  The {\it left panel} has $n=1$, and the {\it right panel} $n=2$. In both panels, disks, squares, diamonds, triangles and inverted triangles have $T/(-\kappa) = 0.029,\, 0.048,\,0.072,\,0.119,\,0.571$, respectively. The {\it left panel} also includes the $T=0$ result from our scaling solution (small red disks).}
\label{fig:magneticfieldhorizon}
\end{figure}

Another quantity of interest is the  magnetic flux density on the horizon, as a function of the proper distance from the rotation axis. Instead of $F_{x\varphi}$, we plot
\be\label{eq:fluxdensity}
\Phi_D \equiv F_{\ell_{\mathcal{H}}\varphi} =  F_{x\varphi}\frac{\dd x}{\dd \ell_{\mathcal{H}}}\,,
\ee
so the  area under the curve directly gives the total flux (up to a factor of $2\pi$ coming from the $\varphi$ integral). The results, for various temperatures, are shown in Fig. \ref{fig:magneticfieldhorizon}.
As expected, the total flux is independent of $T$. Note that the width of the vortex, defined as the region where most of the flux is concentrated, 
is approximately constant as we lower the temperature. This is also expected, since it is essentially the width of the cosmic string when it hits the horizon. Furthermore, the maximum of $\Phi_D$ is a monotonic function  of the temperature, increasing as we decrease $T$. We will see that this last property does not hold for the
 magnetic field  at the conformal boundary at infinity. 

\subsubsection{Field-theoretic  and thermodynamic observables}

We turn now to field-theoretic observables extracted from the asymptotic behavior of our solution. It turns out that there are strong differences between the superconducting case that we are discussing now, and the superfluid case that will be discussed in the next section.

\begin{figure}[h]
\centering
\includegraphics[width=\textwidth]{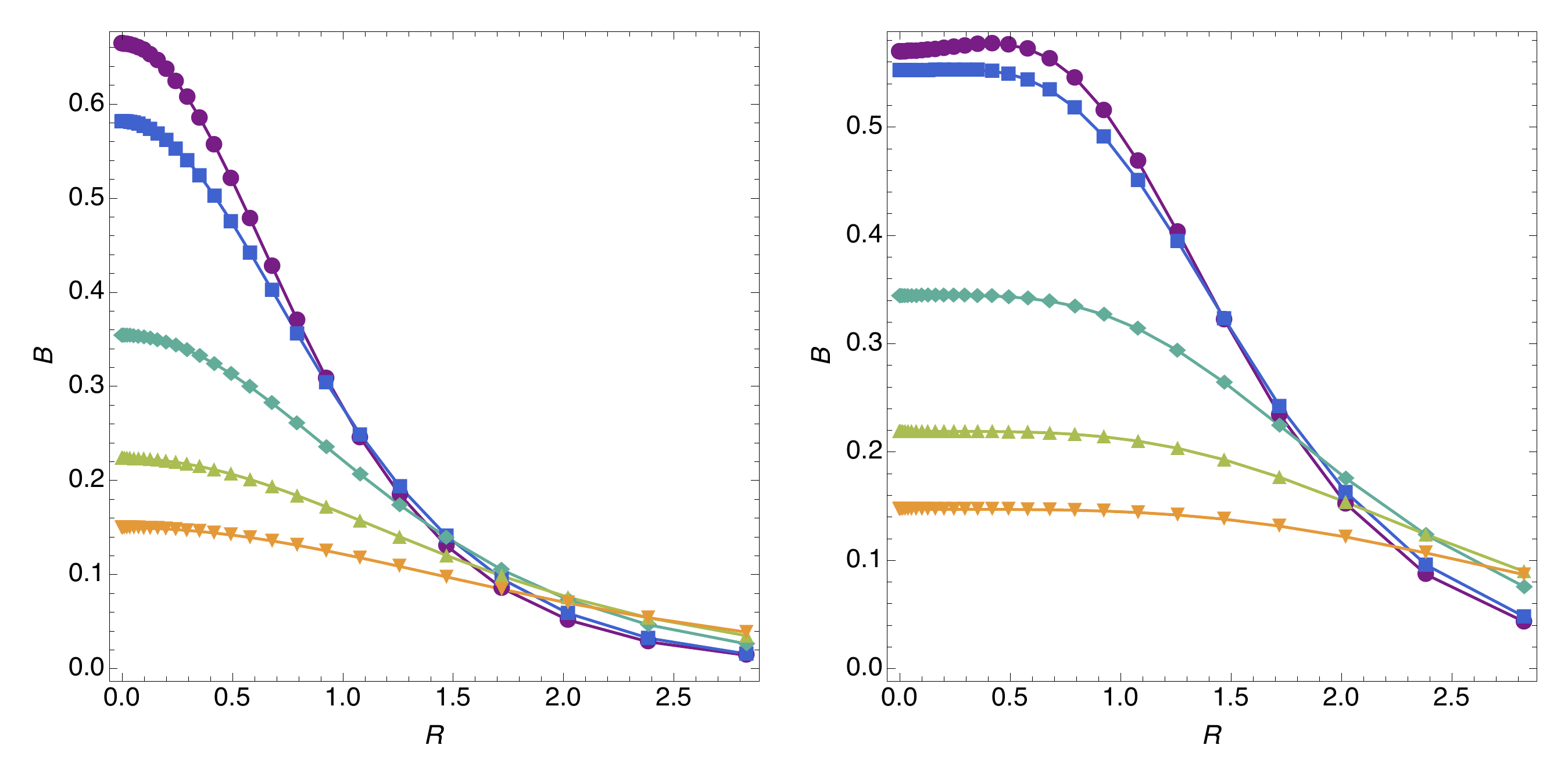}
\caption{Boundary magnetic field profile as a function of $R$, plotted for several values of $T/(-\kappa)$. The {\it left panel} has $n=1$, and the {\it right panel} $n=2$. Here, disks, squares, diamonds, triangles and inverted triangles have $T/(-\kappa) = 0.029,\, 0.370,\,0.495,\,0.546,\,0.571$, respectively.}
\label{fig:magnetic_field_n1_n2}
\end{figure}

We begin with the magnetic field in the boundary theory, $F_{x_1 x_2}\equiv B(x)$.  Here $x_1$ and $x_2$ are boundary cartesian coordinates, defined in the usual way $x_1 = R \cos \varphi$ and $x_2 = R \sin \varphi$. $B(x)$ can be easily expressed as a function of $Q_7$ evaluated at the conformal boundary
\begin{equation}
B(x) = (1-x)^3\left[2 Q_7(x,0)+x \frac{\partial Q_7(x,0)}{\partial x}\right]\,.
\label{eq:magneticfieldjorge}
\end{equation}
In Figs.~\ref{fig:magnetic_field_n1_n2} we plot this boundary magnetic field  for several temperatures and for $n=1,\,2$. Interestingly, its maximum value correlates with the location of the maximum Ricci scalar evaluated along the horizon.
\begin{figure}[h]
\centering
\includegraphics[width=0.5\textwidth]{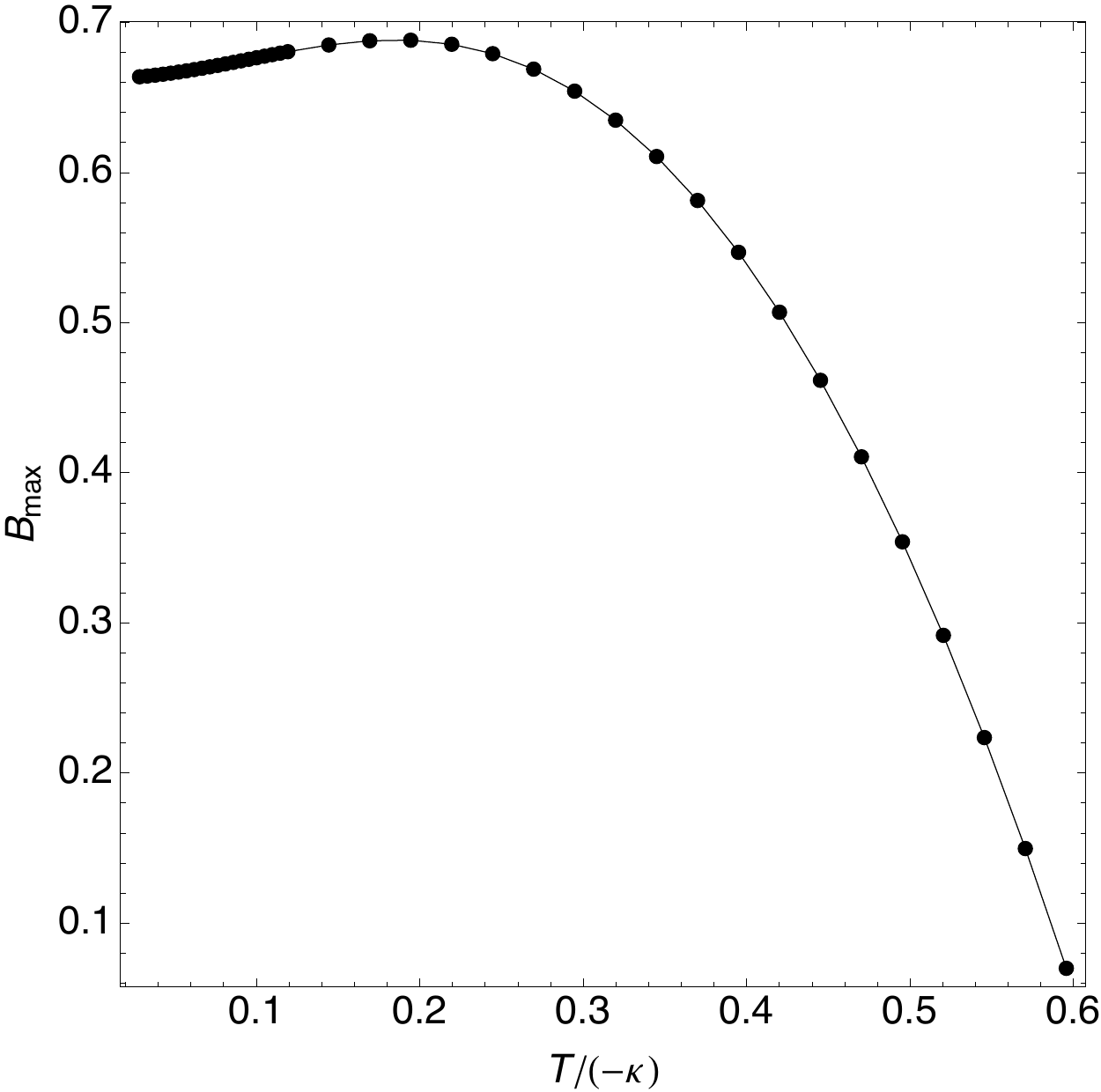}
\caption{Maximum of the boundary magnetic field as a function of $T/(-\kappa)$.}
\label{fig:magneticmaximum}
\end{figure}
It turns out that the maximum magnetic field is not a monotonic function of the temperature. In particular, for $n=1$ and for $T$ smaller than $[T/(-\kappa)]_c\approx 0.185$, the maximum magnetic field starts decreasing with decreasing temperature, see Fig.~\ref{fig:magneticmaximum}. 

\begin{figure}[h]
\centering
\includegraphics[width=0.7\textwidth]{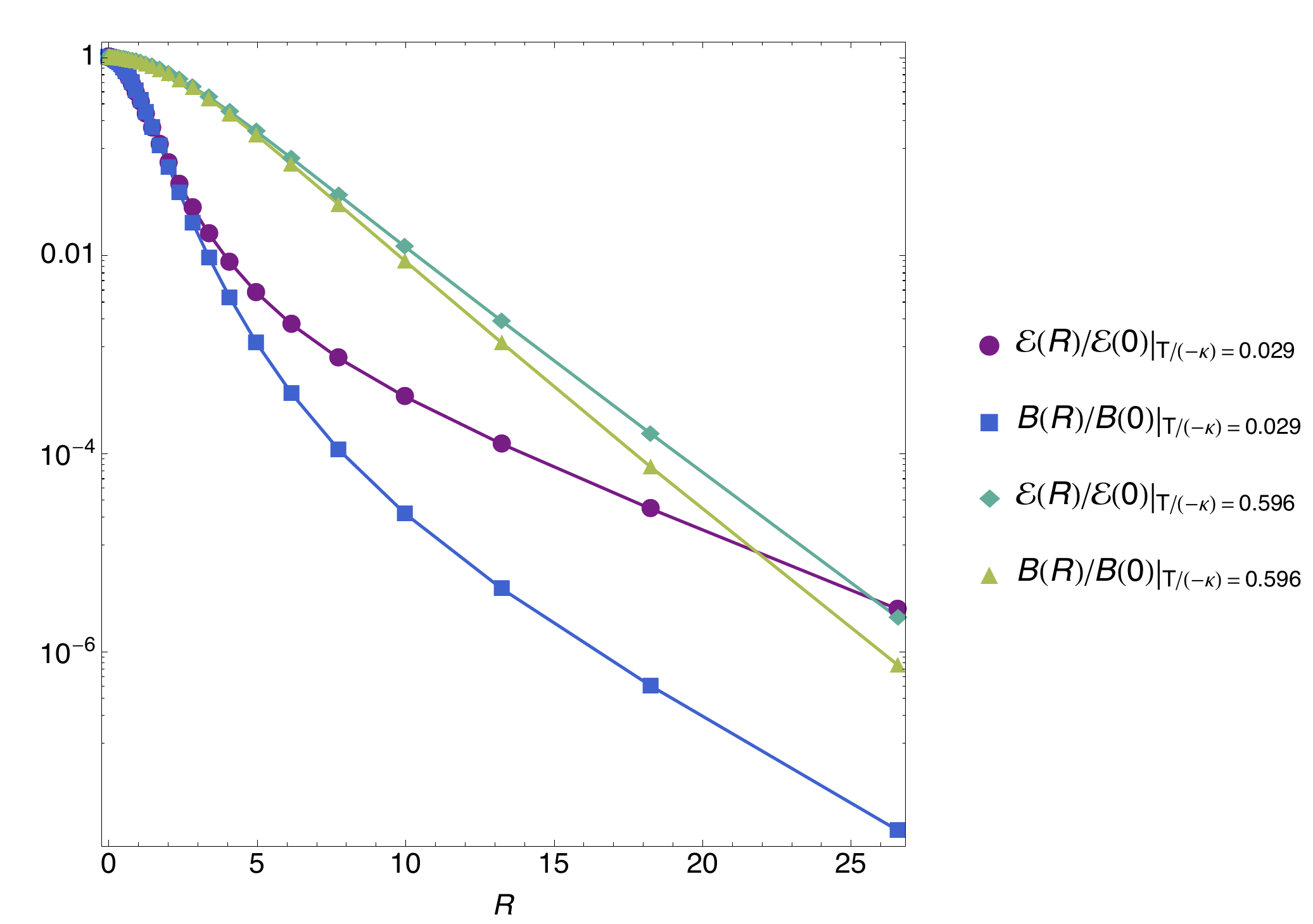}
\caption{Normalized logarithmic plots of energy density and magnetic field as a function of $R$ at two different temperatures. Note that when the temperature is changed the asymptotic slope of the magnetic field changes very little whereas that of the energy density changes significantly. In fact this ``energy screening length'' diverges as $T \to 0$.}
\label{fig:energy_falloff}
\end{figure}

We note that the magnetic field falls off exponentially outside a core radius that is determined by $\ka$. Even as the temperature is taken to zero this core radius remains finite. This should be contrasted with the falloff of the energy density ${\cal E}(R)$, as shown in Fig.~\ref{fig:energy_falloff}. This is well fit at low temperatures by ${\cal E}(R) \sim \frac{e^{-\alpha(T) R}}{R^3}$ but where the inverse ``energy screening length'' $\alpha(T) \to 0$ as $T \to 0$. Thus at precisely zero temperature the vortex sources a long-range disturbance in the stress tensor, due to its interaction with the IR CFT. The exponent of the power law is simply the dimension of the stress tensor. This long-range tail demonstrates a difference between conformal vortices and conventional superconducting vortices, which source no long range fields. The situation is different for superfluid vortices: while presumably the long-range tail discussed here is still present, it will be swamped by a more powerful (and more conventional) IR divergence.

We turn now to the thermodynamics, i.e. entropy, energy and Helmoltz free energy of isolated gravitational vortices at nonzero temperature.\footnote{For each $n$ there is a unique $T=0$ vortex solution with $n$ units of flux. However at $T>0$, the vortex potential acquires temperature corrections which affect the solution. We wish to study this temperature dependence.} We will see that some of their global thermodynamic properties strongly depend on $q$.  We start with the entropy. Since we are working at nonzero temperature, it is clear that the total entropy is infinite, as the black hole horizon extends infinitely far from the vortex. The quantity of most interest is the difference between the entropy with the vortex and the entropy without (at the same temperature). 
 From the above metric, it is easy to see that this  entropy difference is given by
\begin{equation}
\Delta S = \frac{\pi}{2}\int_0^1 \frac{y_+^2\,x}{(1-x)^3}\left[\sqrt{Q_4(x,1)Q_5(x,1)}-\tilde{Q}_4(1)\right]\dd x\,.
\label{eq:difentropy}
\end{equation}
The fact that this expression is finite is another test of the numerics, since one has to cancel a third order pole at $x=1$ in the denominator.

We have actually presented this entropy difference previously in Fig.~\ref{fig:entropydiff}, where we compared it to the impurity entropy of the $T=0$   scaling solution for the $n=1$ vortex. We see that this entropy difference grows as we decrease the temperature, approaching the $T = 0$ result computed previously. This is another illustration of the fact that the vortex causes the horizon to  ``bubble out". The $n=2$ vortex is wider and causes a larger bubble on the horizon with greater area. In fact, the $\Delta S$ computed for $n=2$ is about twice $\Delta S$ computed for $n=1$. We will look at this more closely shortly as it is an indication of whether the $n=2$ vortex can fragment into two $n=1$ vortices. 

The entropy difference remains nonzero at the critical value $[T/(-\kappa)]_c \approx 0.6$. This is the critical temperature for the scalar field to condense \cite{Faulkner:2010gj}, and  beyond this value, the vortex no longer exists.  It might look strange to see the entropy starting at some finite value precisely at $T=T_c$, i.e. the entropy difference seems to be a discontinuous function of the temperature, which seems to be in some tension with the fact that this is a second order phase transition. Importantly, the discontinuity is not in a thermodynamic entropy {\it density} (which would contribute a total entropy scaling with the system size), but rather in a finite impurity entropy that is independent of the system size. Said differently, this comes from the fact we are looking at a single isolated vortex, together with the fact that we are integrating over an infinite domain. The tail extending from $x\sim 0.5$ to $x=1$ is enough to give a finite contribution if we approach $T_c$ from below. We have explicitly checked that this is the case, by truncating the integration to be only over a finite domain, instead of all the way down to $x=1$. If we compute the integral up to \emph{any} $x=x^{\star}<1$, we find that $\Delta S$ is zero at $T=T_c$. Thus the limit of infinite system size and $T \to T_c$ do not commute.

\subsubsection{Superconducting vortex stability}

We now turn to an important physical issue: that of vortex stability. In particular, we would like to study whether an $n = 2$ vortex is unstable to breaking into two $n=1$ vortices. Before presenting our results, we discuss the expectations from the boundary field theory, which will require us to revisit the distinction between Type I and Type II superconductors. 

For a $2+1$ dimensional superconductor, any applied magnetic field will necessarily penetrate the sample\footnote{In a 3+1 dimensional superconductor in an applied field, this is not the case: for a small field, currents running inside the sample can push the field lines {\it around} the boundaries of the sample until a critical field (conventionally called $H_{c1}$) is reached, after which the field will begin to penetrate it. This is clearly not geometrically possible in 2+1 dimensions: in other words, in 2+1 dimensions $H_{c1} = 0$.}. This flux must then locally disrupt the condensate and create some (possibly very small) regions of normal phase. Famously, the way in which this normal phase is distributed is different for Type I and Type II superconductors. Consider the domain wall separating a region of normal phase (carrying magnetic flux) and the superconducting phase. For Type I superconductors this domain wall costs positive energy; thus the system will attempt to minimize the length of this domain wall, which is accomplished by trying to create very large continuous chunks of normal phase that accommodate all the flux, i.e. phase separation. However for Type II superconductors the domain wall costs {\it negative} energy: the system will thus try to {\it maximize} the length of the domain wall by separating the normal phase into as many small pieces as possible. This subdivision will continue until the system hits the quantum limit, with each small piece of normal phase now a vortex carrying a single quantum of flux, arranged in an Abrikosov lattice and preserving superconductivity.

Note that from this behavior we may conclude that for Type I superconductors an $n=2$ vortex should be energetically favored over two $n=1$ vortices, as essentially the vortices attract each other and want to merge together. The opposite is true for Type II: here an $n=2$ vortex wants to break into two $n=1$ vortices, which will repel each other and eventually form an Abrikosov lattice. 

Let us now discuss the microscopic mechanism behind this behavior. The Landau-Ginzburg effective theory of superconductivity has two length scales; the London penetration depth $\lam$ and the coherence length $\xi$. $\lam$ measures how quickly the magnetic field falls off in a superconductor: it is thus the inverse mass of the photon in the Higgsed phase. $\xi$ measures how quickly disturbances of the order parameter fall off: we can view it as the inverse mass of the Higgs boson itself. It is the ratio
\be
\kappa_{LG} = \frac{\lam}{\xi}
\ee
that controls whether the superconductor is Type I or Type II, with $\kappa_{LG} \to 0$ being the Type I limit, and $\kappa_{LG} \to \infty$ the Type II limit. This can be understood by studying the energetics in the vicinity of the domain wall; see e.g. \cite{tinkham} for details. In the framework of Landau-Ginzburg theory, the threshold between the two is at precisely $\kappa_{LG}^{\star} = \frac{1}{\sqrt{2}}$. Thus we conclude that this ratio of correlation lengths should be correlated with vortex stability. 

\begin{figure}[ht]
\centering
\includegraphics[width=0.47\textwidth]{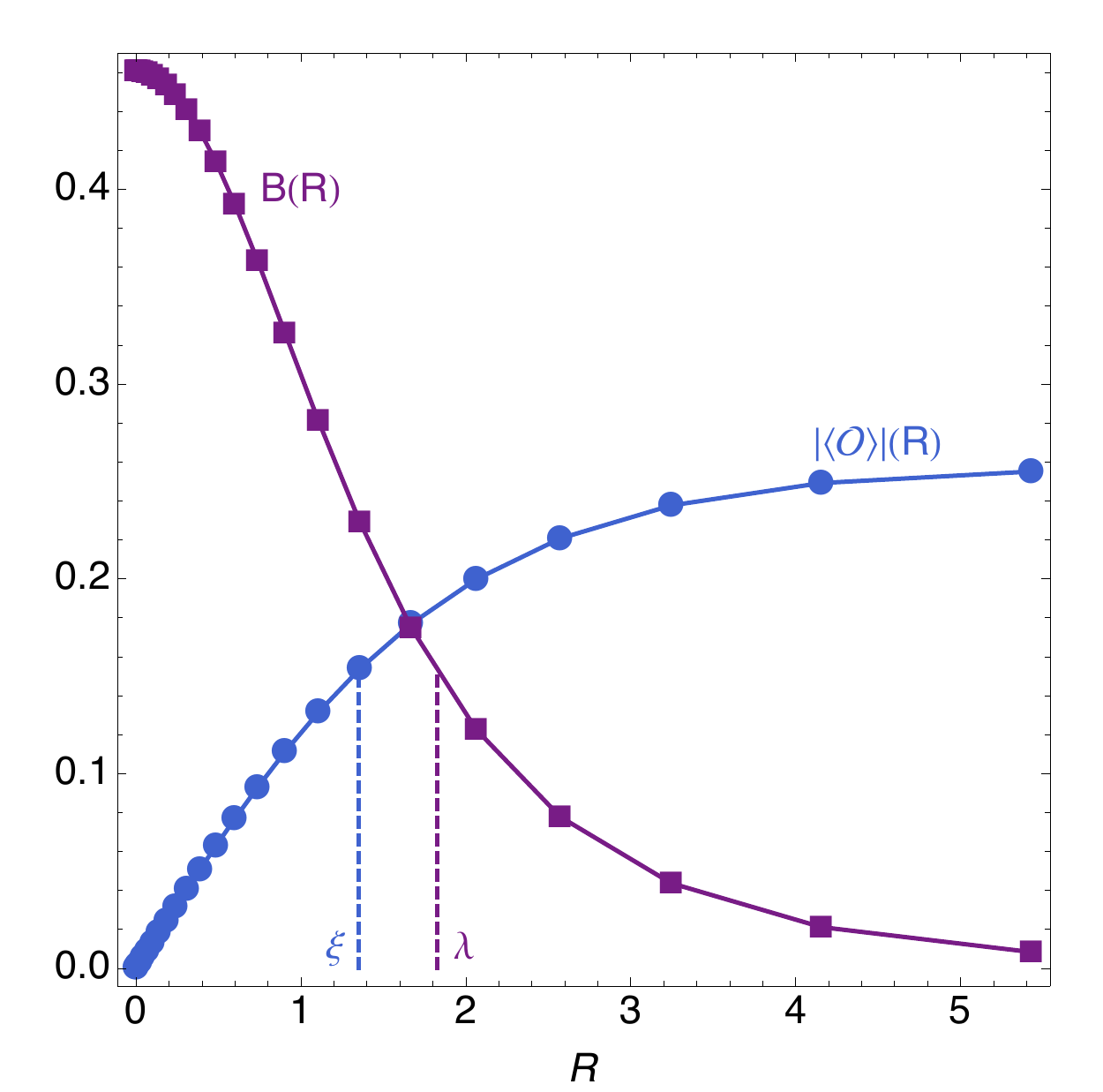} \includegraphics[width=0.47\textwidth]{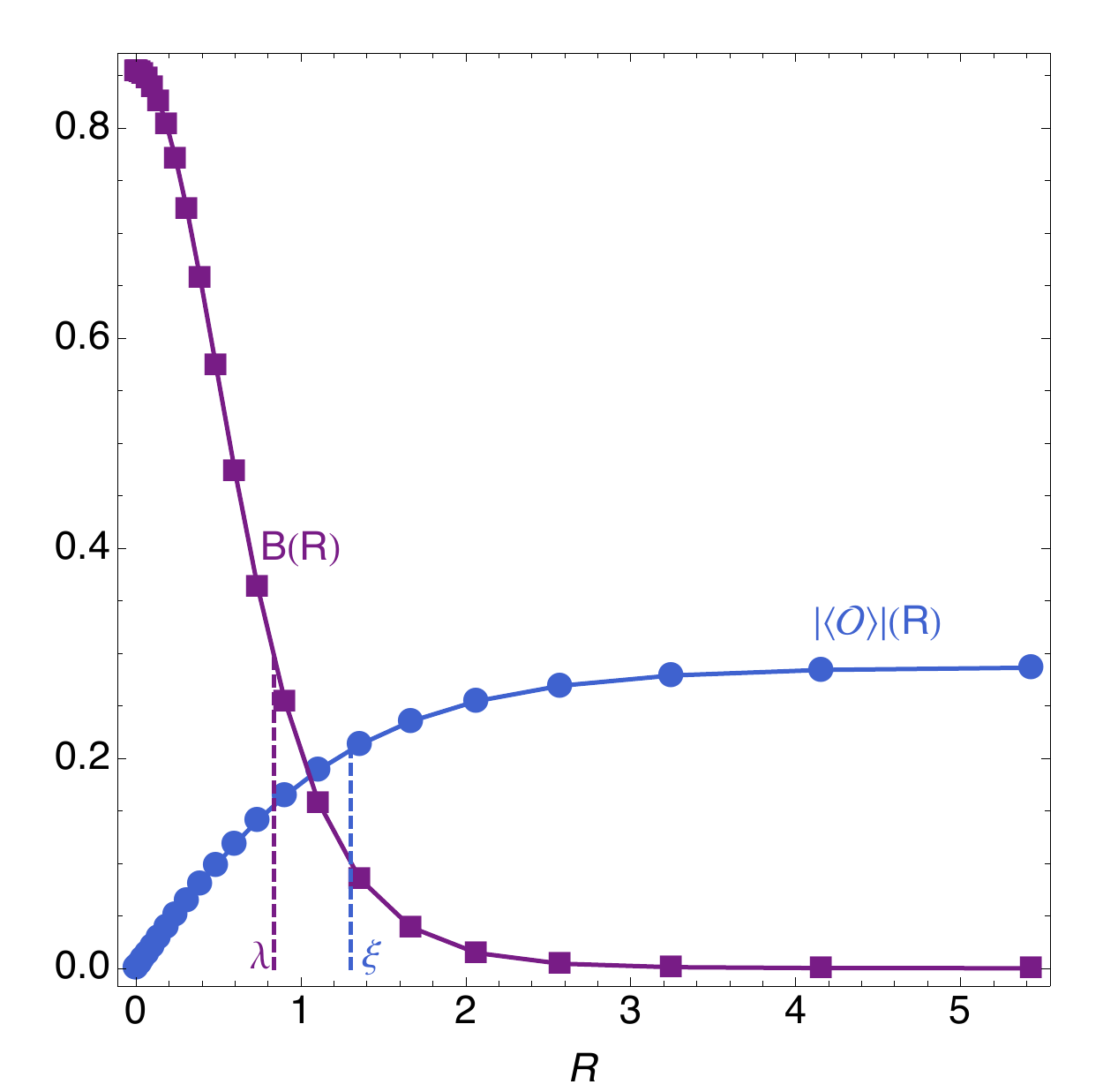}
\caption{Profile of magnetic field $B(R)$ and order parameter $\langle \sO(R) \rangle$ for a single vortex with $q L = 1$ {\it (left panel)} and $q L = 3$ {\it (right panel)}. $\lam$ and $\xi$ are found from exponential fits and measure the rate of fall-off of magnetic field and order parameter, respectively.}
\label{fig:typeII}
\end{figure}

We now return to our gravitational description and see if these expectations are borne out. We will do this for  different values of the scalar charge $q$; interestingly we will find different results. First, we construct the correlation lengths $\lam$ and $\xi$ by fitting an exponential profile (with a subleading power-law correction) to the magnetic field $B(R)$ and the order parameter $\langle \sO(R) \rangle$ for a single vortex:
\begin{equation}
B(R) \sim b \left(\frac{\lambda}{R}\right)^{\alpha}\exp\left(-\frac{R}{\lambda}\right)\,, \qquad \langle \mathcal{O}(\infty) \rangle - \langle \mathcal{O}(R) \rangle \sim o\left(\frac{\xi}{R}\right)^{\beta}\exp\left(-\frac{\sqrt{2} R}{\xi}\right) \ .
\end{equation}\ 

The results are shown in Fig.~\ref{fig:typeII}. It is clear that for $qL =1$ we have $\xi < \lambda$ while for $qL = 3$ we have  $\xi > \lambda$.
The ratio $\kappa_{LG}$ depends weakly on temperature, but for $qL =1$,
$\ka_{LG}  > \frac{1}{\sqrt{2}}$ and we might expect to be firmly in the Type II regime, while for 
$qL = 3$, we have $\ka_{LG} < \frac{1}{\sqrt{2}}$ 
and we expect to be in the Type I regime. For $qL=2$, $\ka_{LG}$ is close to the expected transition at $ \frac{1}{\sqrt{2}}$.

To check these expectations, we compare the entropy at fixed energy and free energy at fixed temperature of an $n=2$ vortex and two $n=1$ vortices.  We will see that both the
 microcanonical and canonical analysis give the same answer for the stability of an $n=2$ vortex.
 
\begin{figure}[ht]
\centering
\includegraphics[width=\textwidth]{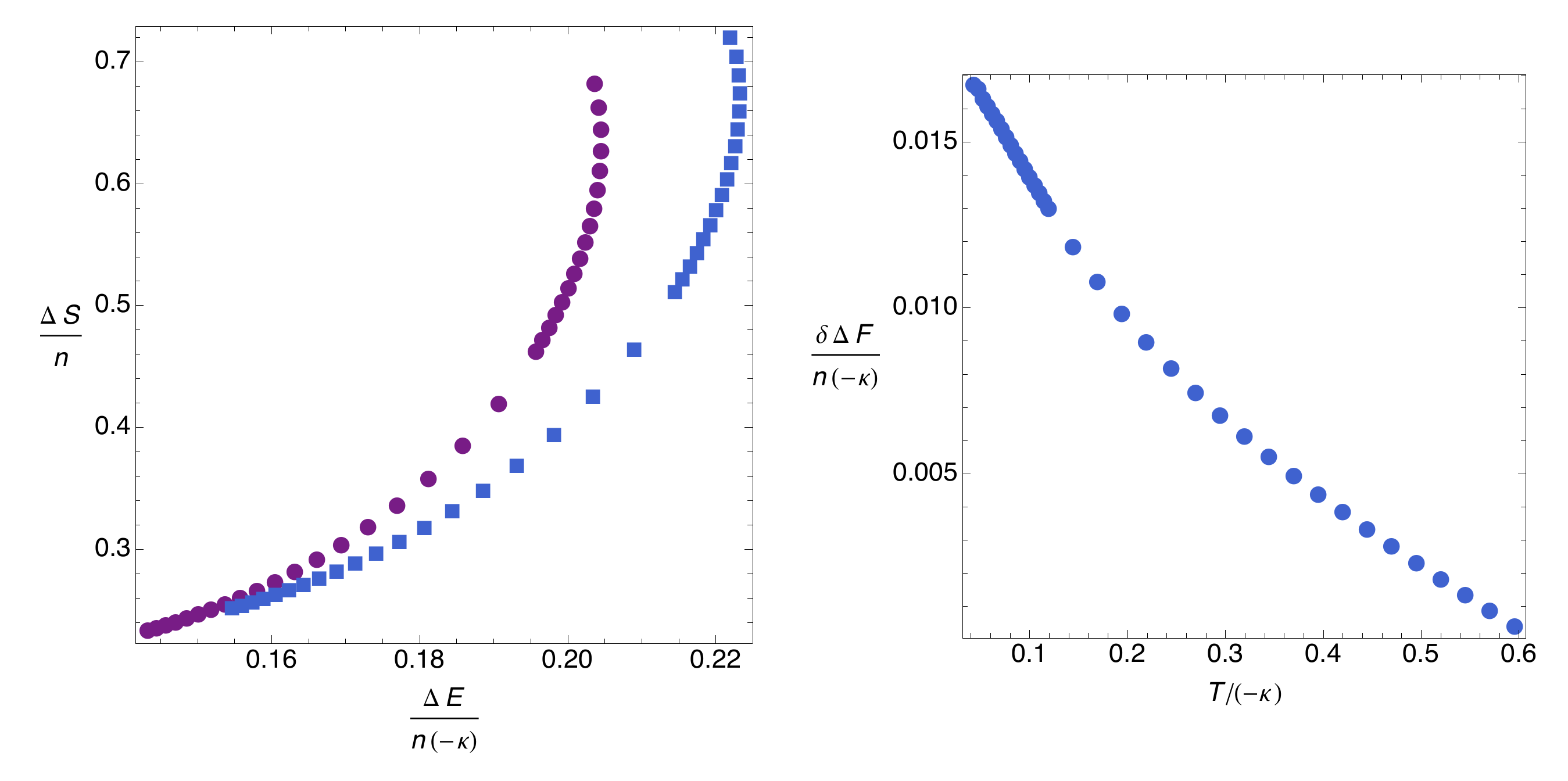}
\caption{For $qL = 1$, the $n=2$ vortex prefers to break up into two $n=1$ vortices. \emph{Left Panel}: entropy difference (\ref{eq:difentropy}) as a function of the energy difference $\Delta E/(-\kappa)$. Disks correspond to $n=1$ and squares to $n=2$. \emph{Right Panel}: the difference in free energies, $\delta \Delta F = \Delta F_{n=2}/2-\Delta F_{n=1}$, as a function of the temperature $T/(-\kappa)$.}
\label{fig:comparison2}
\end{figure}

We start with the $qL =1 $ case. Since the total energy, like the entropy, diverges due to infinite volume, we will work with the difference, $\Delta E$,  which is defined as the difference in energy between the vortex solution and the corresponding HHH black hole at the same temperature. If we use Eq.~(\ref{eq:energystress}), one finds
\begin{multline}
\Delta E = -\frac{y_+^2}{48}\int_0^1 \frac{x}{(1-x)^3}\Bigg\{3 y_+\left(\left.\frac{\partial^3 Q_1}{\partial y^3}\right|_{y=0}-\left.\frac{\partial^3 \tilde{Q}_1}{\partial y^3}\right|_{y=0}\right)\\+19\kappa \left[x^n Q_6(x,0)^2-\tilde{Q}_4(0)^2\right]\Bigg\}\,\dd x\,.
\label{eq:energydiffQ}
\end{multline}
Like the entropy, the fact that this expression is finite is in itself a test of the numerics.

First, we will plot the entropy difference (\ref{eq:difentropy}) as a function of the energy difference (\ref{eq:energydiffQ})   for both $n=1$ and $n=2$. This comparison is appropriate for a  microcanonical ensemble; the solution with the higher entropy will dominate. The results are illustrated on the {\it left panel} of Fig.~\ref{fig:comparison2} for $q\,L=1$. We have also divided $\Delta S$ by the respective value of $n$, since we want to compare the entropy of two isolated $n=1$ vortices with the entropy of a single vortex with $n=2$. Note that the $n=2$ vortex appears to be always unstable to breaking into two $n=1$ vortices. The same result is obtained in a canonical ensemble when we compare the free energies $F = E -TS$.  The {\it right panel} of Fig~\ref{fig:comparison2} shows a plot of $\delta \Delta F = \Delta F_{n=2}/2-\Delta F_{n=1}$. The fact that this quantity is always 
positive confirms that the $n=2$ configuration is always unstable towards breaking into two $n=1$ vortices. Thus this holographic superconductor is Type II. This is in agreement with our study of the Landau-Ginzburg parameter $\ka_{LG}$ above. 

\begin{figure}[ht]
\centering
\includegraphics[width=\textwidth]{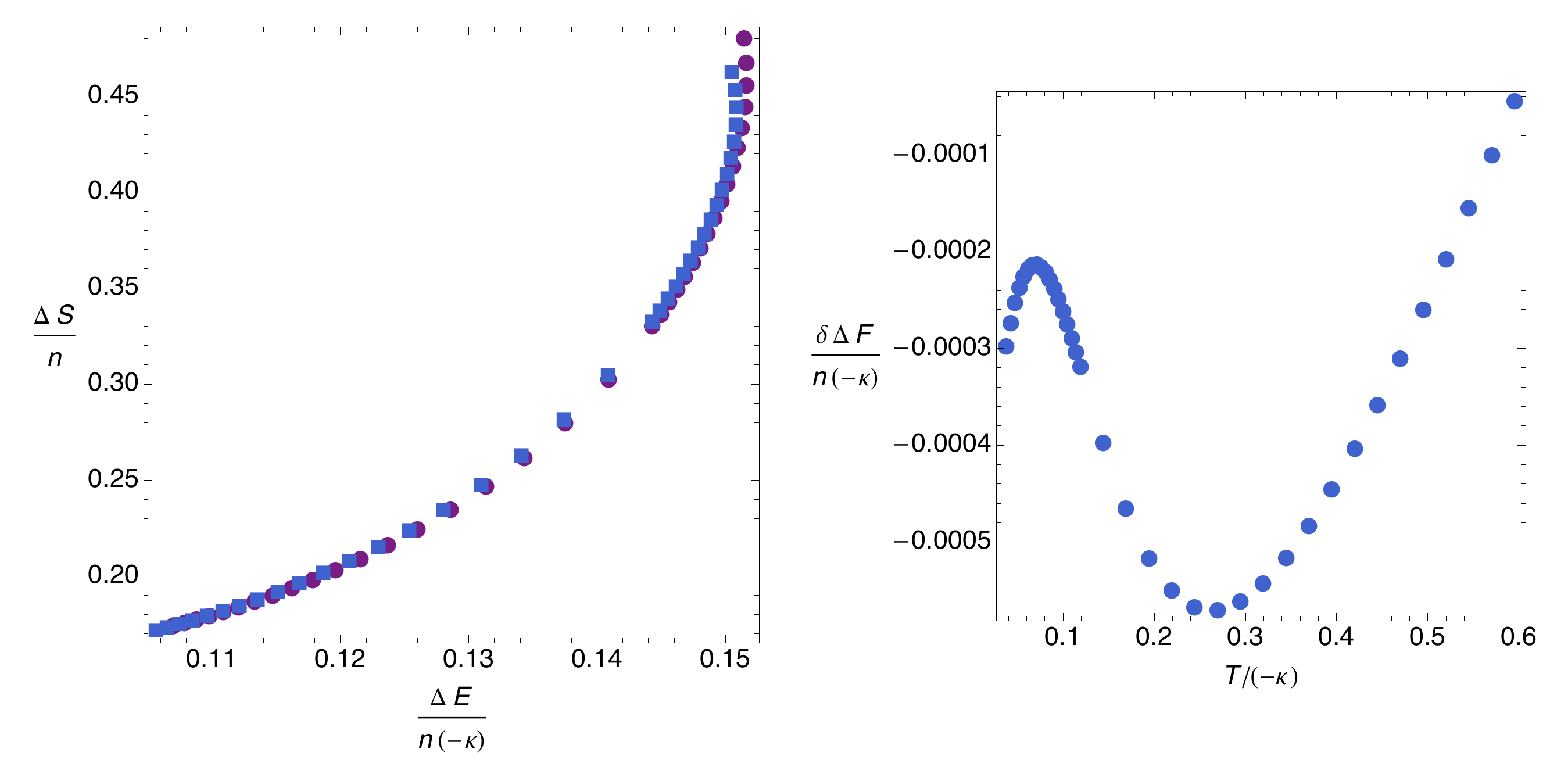}
\caption{For $qL = 2$, the $n=2$ vortex is (slightly) favored over two $n=1$ vortices. \emph{Left Panel}: entropy difference (\ref{eq:difentropy}) as a function of $\Delta E/(-\kappa)$ for $q\,L =2$. Disks correspond to $n=1$ and squares to $n=2$. \emph{Right Panel}: the difference in free energies, $\delta \Delta F = \Delta F_{n=2}/2-\Delta F_{n=1}$, as a function of $T/(-\kappa)$.}
\label{fig:comparison}
\end{figure}

We now repeat the analysis for  $q L = 2$, shown in Fig.~\ref{fig:comparison}: things have changed, and now the entropies and free energies of the two configurations are very similar. Although the points are very close, we have checked that in both cases the $n=2$ vortex is {\it favored} over two $n=1$ vortices. Thus for this value of the scalar charge the vortex is Type I, but is very close to the threshold for the crossover to Type II. This agrees perfectly with our expectations from studying $\ka_{LG}$, which for this value of the charge was very close to the critical value $\frac{1}{\sqrt{2}}$. Finally, for $qL = 3$, we have verified that both the entropy and free energy differences are larger than $qL = 2$, and continue to favor the $n=2$ vortex over the two $n=1$ vortices. This again agrees with our expectations from studying $\ka_{LG}$.

We end our discussion of vortex stability with a final comment: we have seen that vortex stability is precisely the distinction between Type I and Type II superconductors. From our analysis it is clear that whether or not a particular holographic superconductor is Type I or Type II depends on the detailed dynamics, i.e. the non-universal ratio of two different correlation lengths, which appears to be sensitive to (for example) the precise value of the scalar charge. 
While  most of the literature on  holographic superconductors states that they are Type II  \cite{Hartnoll:2008kx,Domenech:2010nf}, this was originally based on the  fact that the scalar condensate starts to condense at a nonzero value of the magnetic field. This was interpreted as $B_{c2}$, the value of the magnetic field in a Type II superconductor below which vortices penetrate the superconductor without destroying it completely. We have checked that in all our examples, the scalar condensate starts to condense at a nonzero value of the magnetic field. So it is now clear that this is not a sufficient condition to determine the type of  superconductor (it could simply indicate phase separation in a Type I superconductor.) Furthermore, studying only a single vortex does not provide enough information to settle this question. One must perform a more detailed comparison of free energies of the sort performed here, and indeed over a wide parameter range we have seen that it is possible for a holographic superconductor to be Type I. In the conclusion we discuss some directions to investigate this further.


\subsection{Superfluid vortices}

We now turn from superconducting vortices to superfluid ones. Recall that the superfluid vortex differs from the superconducting vortex in that the latter is sourced by a boundary magnetic field while the former has no applied field, but does possess a boundary current $J_{\varphi}$. Thus they differ at the conformal boundary but have the same boundary conditions at the horizon. It is then no surprise that quantities measured at the horizon behave similarly in the two phases. In particular, all of the observables studied in Section \ref{sec:sc_hor} -- involving properties of the superconducting horizon -- are largely the same, and we will not discuss them further.

\begin{figure}[ht]
\centering
\includegraphics[width=\textwidth]{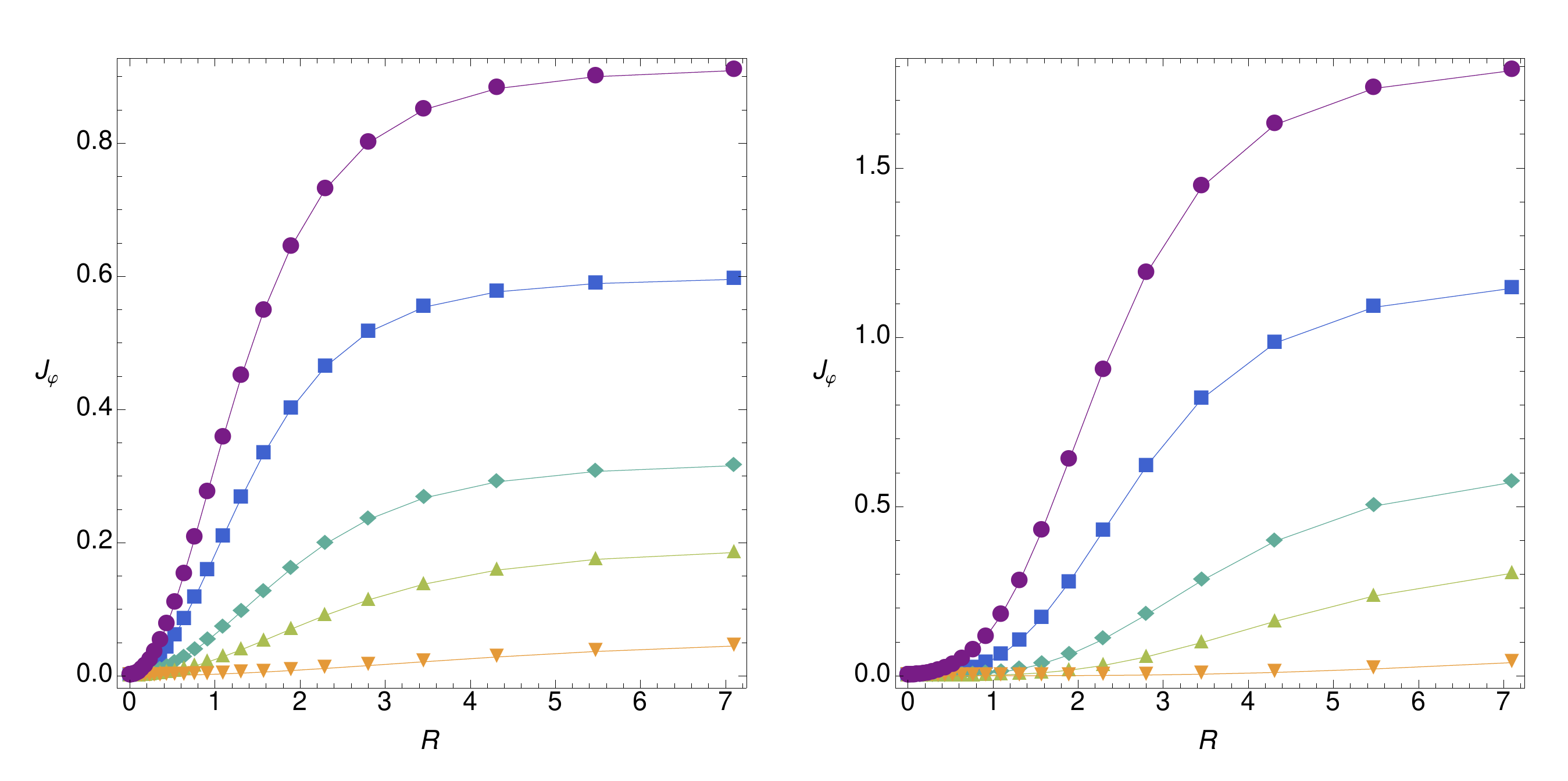}
\caption{Boundary  current profile of a superfluid vortex as a function of $R$, plotted for several values of $T/(-\kappa)$. The {\it left panel} has $n=1$, and the {\it right panel} $n=2$ (both are for $q L=2$). Here, disks, squares, diamonds, triangles and inverted triangles have $T/(-\kappa) = 0.029,\, 0.370,\,0.495,\,0.546,\,0.571$, respectively.}
\label{fig:SF:magnetic_current}
\end{figure}

\begin{figure}[h]
\centering
\includegraphics[width=0.5\textwidth]{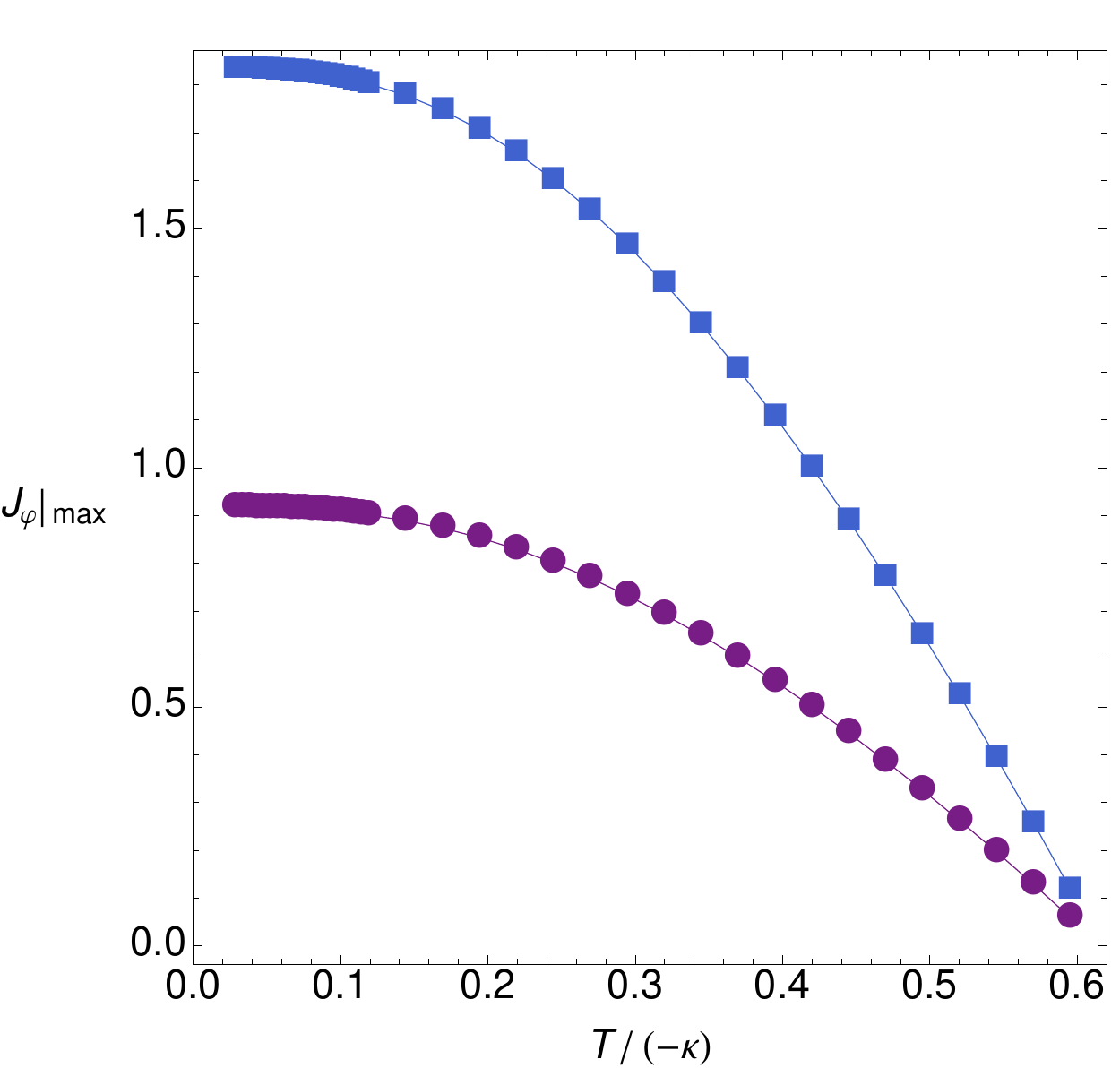}
\caption{Maximum of the boundary  current as a function of the $T/(-\kappa)$. Here, disks and squares describe the superfluid vortex phase with $n=1$ and $n=2$, respectively (both for $qL=2$).}
\label{fig:SF:magJmax}
\end{figure}

We now turn our attention to physical properties that are unique to the superfluid vortices.
The boundary current can be extracted from the bulk fields as:
\begin{equation}
J_\varphi(x) = y_+ \,x^2 \,\frac{\partial Q_7(x,0)}{\partial y}\,. 
\label{eqG:current}
\end{equation}
Fig.~\ref{fig:SF:magnetic_current} shows the profile of the boundary  current 
$J_\varphi$ as a function of the boundary radius $R$, for several values of temperature (the {\it left panel} is for $n=1$ while the {\it right panel} is for $n=2$).  
We see that the current $J_\varphi$ vanishes at the origin ($R=0$) of the superfluid vortex and then, as one moves away from the vortex core, it increases monotonically, initially with a big slope and then flattening out as $R\to +\infty$ to become a constant. 
As expected, the highest values of the current are attained far away from the core of the vortex and this maximum value decreases as the temperature of the system increases, as better illustrated in Fig.~\ref{fig:SF:magJmax}. Increasing $n$ increases the net circulation and the maximum value of the current, as shown in Figs.~\ref{fig:SF:magnetic_current} and \ref{fig:SF:magJmax}. These plots are for $qL=2$. Not shown in these plots is the fact that, for a given temperature $T/(-\kappa)$ and winding number $n$, we find that $J_\varphi{\bigl |}_{\rm max}$ increases as $qL$ grows.

\begin{figure}[h]
\centering
\includegraphics[width=\textwidth]{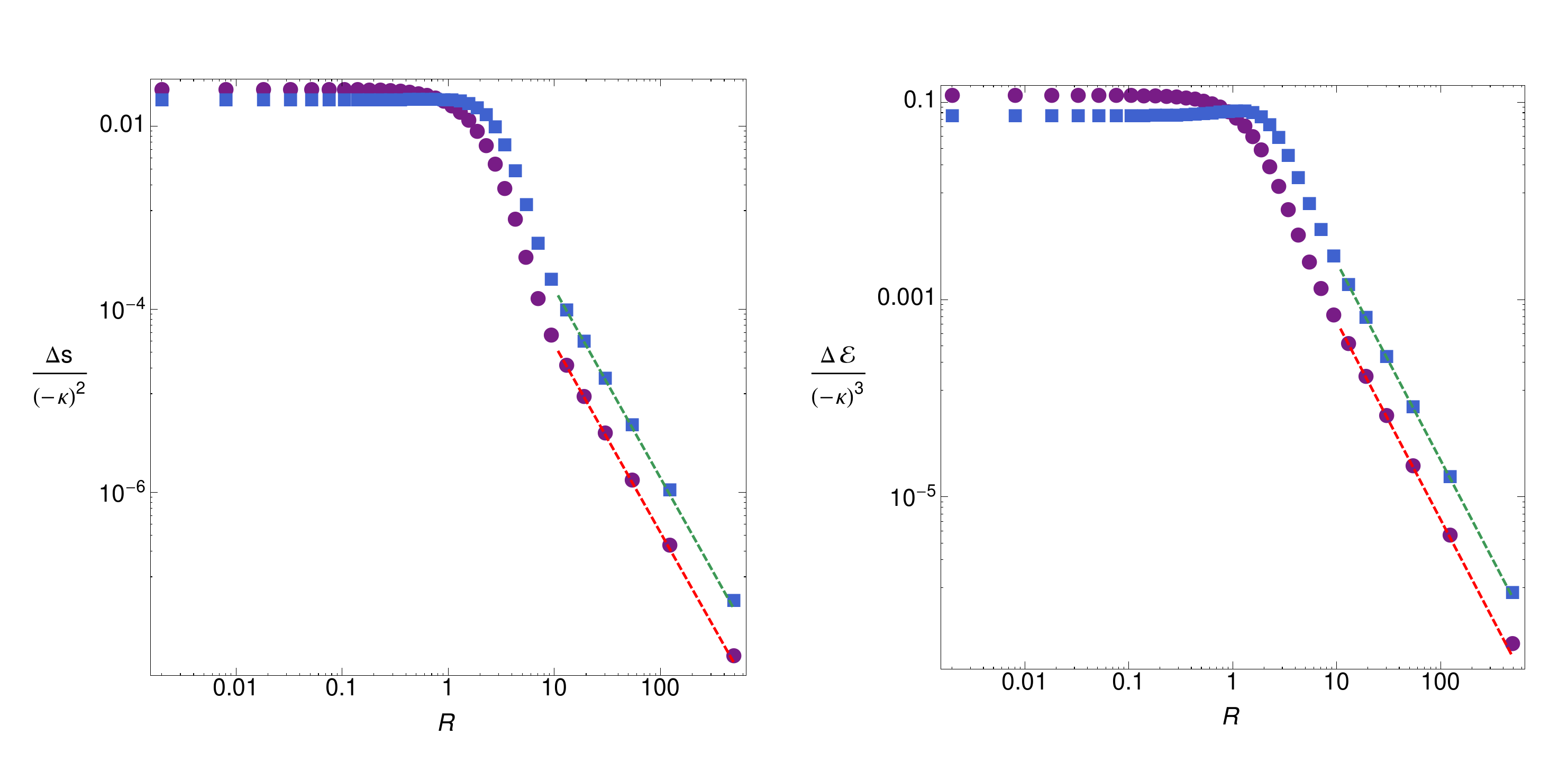}
\caption{Entropy density ({\it left panel}) and energy density ({\it right panel}) as a function of $R$ for the vortex superfluid phase. Here, disks and squares describe, respectively, isolated vortices with $n=1$ and $n=2$ (both for $qL=2$). At large $R$, both densities decay polynomially as $1/R^2$ as described by the dashed curves that give the best fit of the asymptotic tails. For example, for $n=1$ one finds the fit $\Delta s/\left(-\kappa \right)^2=A_0/R^{\alpha }$ with 
$\{\alpha \sim 2.006\pm 0.001,\:  A_0\sim 0.0040\pm 0.0001\}$ and $\Delta {\mathcal{E}}/\left(-\kappa \right)^3=B_0/R^\beta$ with $\{ \beta \sim 2.005\pm 0.001,\: B_0\sim  0.0602\pm 0.0002\}$.}
\label{fig:SF:SEdensities}
\end{figure}

We now turn to the thermodynamics, i.e. the entropy, energy and Helmoltz free energy. Before discussing our gravitational results, we briefly recall the expectations from field theory. A vortex in a conventional superfluid has an energy that logarithmically diverges with the system size. Recall that the low-energy dynamics of a superfluid is given by the action for a Goldstone mode $\th$:
\be
S = \rho_s \int d^3 x\;(\nabla \th)^2 \ . \label{goldac}
\ee
A vortex with charge $n$ has $\th(R \to \infty) \sim n \varphi$ with $\varphi$ the azimuthal angle around the vortex. Evaluating the energy following from \eqref{goldac} on such a configuration, we find 
\be
E \sim  \rho_s \int dR \frac{1}{R} n^2 \sim \rho_s n^2 \log\le(\frac{R_{\max}}{a_0}\ri), \label{enIRdiv}
\ee
where $a_0$ is the vortex core size and $R_{\max}$ an IR cutoff. This is a standard result. Perhaps slightly less obvious is the fact that the first law of thermodynamics $dE = T dS$ states that at finite temperature this IR divergent energy implies also an IR divergent {\it entropy}. One way to understand this is to note that at finite temperature the current $J_{\varphi}$ will contain a normal component, which falls off slowly in space and carries an associated thermal entropy.

We now return to our gravitational description and compute the bulk energy density difference $\Delta {\cal E}$ \eqref{eq:energydiffQ} and entropy density difference $\Delta s$ \eqref{eq:difentropy} from our bulk gravitational solution. As expected from the discussion above, both of these quantities decay only as $R^{-2}\sim(1-x)^2$ at large boundary radius $R$, as shown in Fig.~\ref{fig:SF:SEdensities}. The volume integrals of both these densities diverge logarithmically at large $R$, as expected.\footnote{Recall that in the superconducting phase $\Delta S$ and $\Delta E$ are finite because the corresponding densities  $\Delta s$ and $\Delta {\cal E}$ have an asymptotic exponential decay.} 
 The entropy density difference has a temperature dependent coefficient,  $\Delta s \sim f(T) n^2 /R^2$, and we have verified that the coefficient $f(T)$ vanishes as $T \to 0$. This is expected: it is the thermally excited normal component of the current that is contributing to the IR divergence, and this entropy should indeed vanish as $T \to 0$.\footnote{Note that it is not obvious that the normal component itself should vanish -- defining this precisely in a holographic superfluid is tricky, but there are indications from \cite{Horowitz:2013jaa} that this normal component does not vanish at $T = 0$. We are simply stating that the entropy that it carries vanishes.} At precisely $T = 0$ we expect the entropy of the superfluid vortex to be equal to that of the superconducting vortex, with both answers equal to the impurity entropy arising from the scaling solution, but the IR divergence makes it very difficult to check the approach to this limit.

\begin{figure}[ht]
\centering
\includegraphics[width=0.5\textwidth]{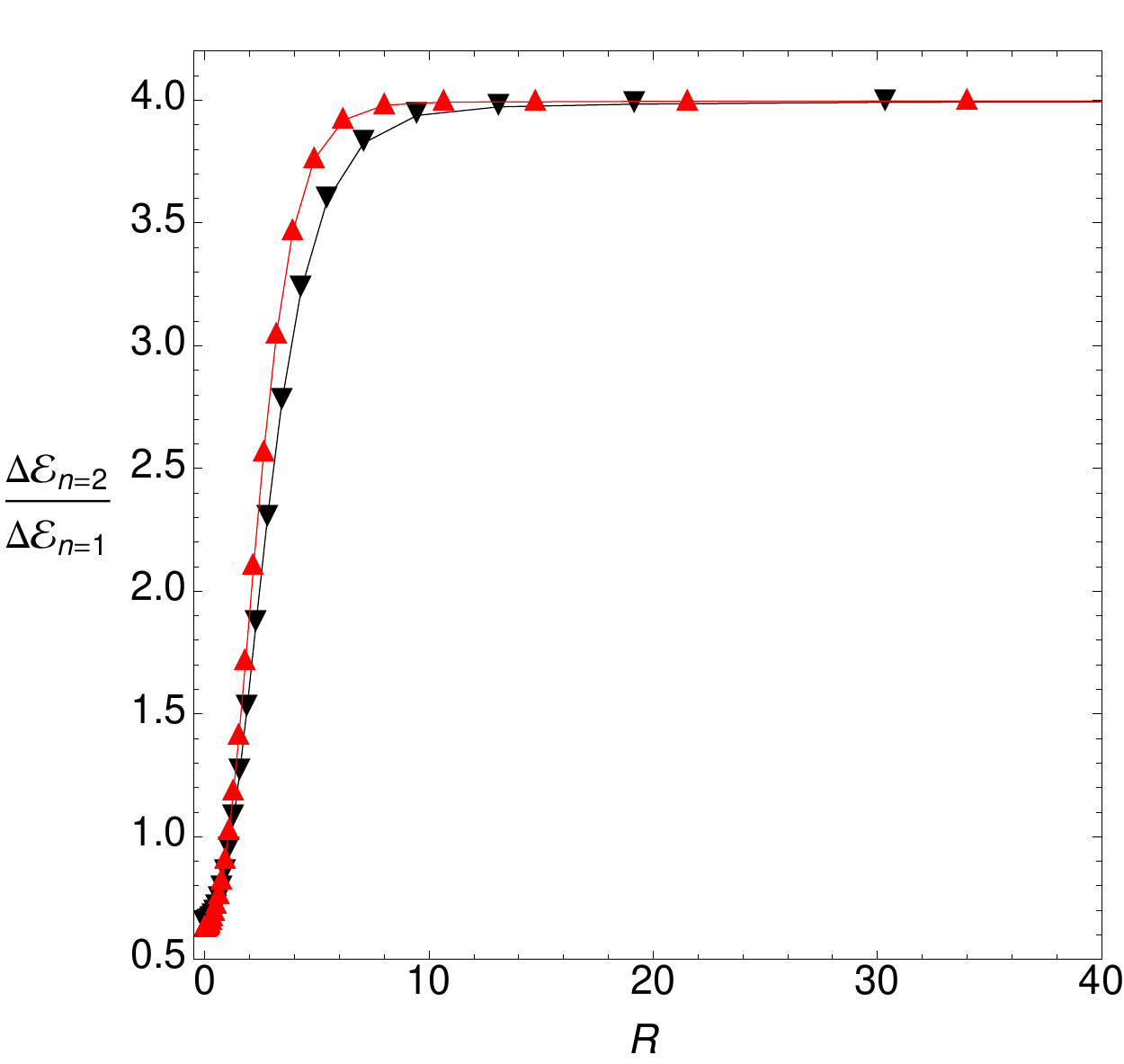}
\caption{Ratio of local energy density of $n=2$ vortex to $n=1$ vortex at $T/(-\ka) = 0.12$. Red triangles indicate $qL = 2$, black inverted triangles $qL = 1$. As expected, at long distances an IR divergence of the form \eqref{enIRdiv} dominates, scaling with the vortex charge as $n^2$.}
\label{fig:SF:ratioenerg}
\end{figure}

We have checked that the coefficient of the IR divergences depend on the vortex winding charge as expected from \eqref{enIRdiv} (see e.g. Fig.~\ref{fig:SF:ratioenerg}). 
Note that these IR divergences make the question of vortex stability somewhat different in a superfluid as opposed to a superconductor: as both the energy and entropy are dominated by the IR divergence which scales with the vortex charge as $n^2$, one concludes that any high-charge vortex should want to dissociate into vortices with the minimal charge $n = 1$, which will then feel a mutual long-range repulsive force, independent of the details of the dynamics. This stability result is in line with the time evolution study done in \cite{Adams:2012pj}, where (in a different setup) it was found that holographic superfluid vortices with high winding charge, introduced in the system as initial data,  rapidly decay into $n=1$ superfluid vortices.

\section{Forces on a moving vortex from conformal invariance} \label{sec:confforce}

We have presented a detailed discussion of the properties of a vortex in a holographic superfluid/superconductor; as we emphasized at various points, many of the facts that we report can be usefully organized by realizing that at low energies the vortex can be viewed as a conformal defect, with a CFT$_1$ living on it. In this section we switch gears and use the defect conformal invariance to compute the forces on a moving vortex in terms of universal data. In particular, we show that there exist Kubo formulas for these forces in terms of defect-localized operators. This section does not use our gravitational description in any way, and should apply to any situation where a vortex coexists with conformal invariance. 

As described before, the vortex worldline hosts a CFT$_1$, which may be characterized by the spectrum of operators living on the defect. The full spectrum of operators depends on the theory in question, but every defect has at least a {\it displacement operator} $D^i$. Adding $D^i$ to the full CFT action corresponds to shifting the location of the defect. It is thus intimately related to the breaking of translational symmetry; on $\mathbb{R}^{2,1}$ the following Ward identity is satisfied:
\be
\p_{\mu} T^{\mu i} = D^i \delta^{(2)}(x)\,. \label{wardident}
\ee
Note that this relation fixes the dimension of $D^i$ to be $2$, and the correlation function of $D^i$ then takes the form
\be
\langle D^i(t) D^j(0) \rangle = \frac{C_D \delta^{ij}}{t^4} \,.\label{ddcorr}
\ee
As \eqref{wardident} fixes the normalization of $D^i$, $C_D$ is a meaningful and universal number characterizing the defect. 

Now in general, if a vortex with circulation $ \hat{\ka}$ in any superfluid is moved through the medium at finite temperature with velocity $v$, it will experience a force whose most general form is
\be
F_i = \hat{\ka} \le(-\ga v_i + \rho_M \ep_{ij} v^j\ri) \equiv \sigma_{ij} v^j \,.
\ee
There are two components: a diagonal frictional force parametrized by $\gamma $ and a transverse force -- called the Magnus and/or Iordanskii force -- parametrized by $\rho_M$. The precise nature of these forces in a conventional superfluid is a matter of some controversy: in particular the coefficient $\rho_M$ is thought to be related to a combination of the superfluid and normal fluid densities, but the precise combination remains somewhat uncertain, with different arguments giving different results \cite{PhysRevB.55.485,PhysRevLett.79.1321,PhysRevLett.76.3758}. In our case both of these densities are zero and the transverse force identically vanishes, so we will have little to say about this. These forces were computed for a holographic superfluid vortex in a probe limit at high temperature in \cite{Iqbal:2011bf}. In this work we will study the opposite low-temperature limit.  

In the defect CFT formalism there is an elegant expression for these forces. Consider a general defect moving with velocity $v^i$. This corresponds to deforming the CFT by the displacement operator $D^i$ with a time-dependent coefficient:
\be
\delta S_{CFT} = \int dt D^i(t) (v_i t) \ .\label{defst}
\ee
We would now like to calculate the force on the vortex. This force is simply the non-conservation of the stress-tensor in the presence of the moving vortex, and is easily found from the Ward identity \eqref{wardident}
\be
F^i = \langle \p_t T^{ti} \rangle_v = \langle D^i \rangle_v \delta^{(2)}(x^i - v^i t) \ . 
\ee
Thus we simply need to compute the expectation value $\langle D^i \rangle_v$ in the deformed state given by \eqref{defst}. This is a problem in linear response; to lowest order in $v$ the answer is simply given by the retarded correlator of $D^i$, which we can express in frequency space as
\be
\langle D^i(\om) \rangle_v = \frac{\langle D^i(\om) D^j(-\om) \rangle}{i\om} \, v_j \ .
\ee
Thus we have identified a Kubo formula for the force tensor $\sigma_{ij}$:
\be\label{eq:twoD}
\sigma^{ij} = \lim_{\om \to 0} \frac{\langle D^i(\om) D^j(-\om) \rangle}{i\om}, \ .
\ee
where it is understood that we are evaluating a retarded correlator. This is one of the main results of this section. 

We now turn to the computation of this two-point function. The dimension of $D^i$ is fixed to be $2$, and at zero temperature we have 
\be
\langle D^i(\om) D^j(-\om) \rangle_{T=0} =  \frac{C_D\pi}{3} \,\delta^{ij}(-i\om)^3\,.
\ee
The overall prefactor is obtained from Fourier transformation of the position-space correlator \eqref{ddcorr}. The $\om \to 0$ limit of this vanishes, as expected: at zero temperature the CFT state is Lorentz invariant, and so a vortex moving at constant speed does not know it is moving. 

At finite temperature $T \neq 0$ the situation is different. Generically at finite temperature we expect nontrivial spectral weight as $\om \to 0$, i.e. if we expand the answer in powers of $\om$ we expect an answer of the form:
\be\label{eq:twoDT}
\lim_{\om \to 0}\langle D^i(\om) D^j(-\om) \rangle_{T} = \frac{\tilde{C}_D \pi}{3}(2\pi T)^2 (i\om) \delta^{ij}+ \sO(\om^2), \ .
\ee
where $\tilde{C}_D$ is a coefficient that we expect to be related to $C_D$ and which depends on the theory in question. 

Finite temperature correlators of CFT$_1$ operators respecting $SL(2,\mathbb{R})$ invariance have been previously calculated in \cite{Faulkner:2011tm,Faulkner:2009wj}. Those results are entirely fixed by conformal invariance: there is a transformation of the time coordinate that can be used to place the
$T=0$ CFT$_1$ at finite temperature, and the full finite $T$ correlator can be obtained from the known conformal transformation of the vacuum CFT$_1$ operators. However that transformation has a nontrivial action on the fields outside of the defect, placing them in a different state that is not obviously equivalent to the thermal state, and thus those results do not appear to immediately apply. It would be useful to understand if that formalism could be extended to this case; this would allow an explicit calculation of $\tilde{C}_D$ in terms of $C_D$.

From \eqref{eq:twoD} and \eqref{eq:twoDT},  the force tensor $\sigma^{ij}$ is simply
\be
\sigma^{ij} \equiv \sigma \delta^{ij} = \frac{4 \pi^3}{3}\tilde{C}_D T^2 \delta^{ij}  \ .  \label{fric}
\ee
This force represents the frictional drag on the vortex as we drag it through the excited medium. 

It is interesting to compare this to the drag force on a moving vortex in an ordinary (non-holographic) superfluid, which as we argued earlier is essentially empty of excitations at low energies. At a temperature $T$ there is a gas of thermally excited Goldstone modes. By scale invariance the momentum density perceived by the moving vortex from these modes is $\langle T^{0i} \rangle \sim T^3 v$. Each of these modes has a cross section $\sigma \sim T a_0^2$ for interaction with the vortex, where $a_0$ is the radius of the core \cite{PhysRev.136.A1488,PhysRevB.55.485}. Thus the force in a conventional superfluid is $\sigma \sim T^4 a_0^2$, a higher power of $T$ than that arising from \eqref{fric}. This extra suppression is due to the existence of the UV scale $a_0$ in the answer: unlike the pure conformal answer \eqref{fric}, which contains no other scales, this frictional force arises from a leading ``irrelevant'' deformation to an otherwise empty theory. 

We note that the knowledge of this force lets us trivially compute the Nernst effect arising from a dilute gas of these vortices. We briefly review the physics of the Nernst effect; consider taking a superconductor and applying a magnetic field $B$ into the sample together with a temperature gradient along the $x$ direction. In general this will set up an electric field $\vec{E}$ perpendicular to the temperature gradient; the Nernst signal is defined to be
\be
e_N = \frac{E}{|\nabla T|}\,.
\ee
We now compute this in our setup. The magnetic field can only be carried by vortices, each of which carries a flux $\frac{2\pi}{q}$; thus we find a vortex density $n = \frac{q B}{2\pi}$. Furthermore in a temperature gradient each vortex will feel an entropic force, arising from the fact that it has an intrinsic entropy:
\be
F^i_{thermal} = s_{tr} \p_i T\,.
\ee 
In general the coefficient $s_{tr}$ is called the ``transport entropy''. In the conformal setting it would be very interesting to understand the precise relation between this thermal entropy and the defect entropy defined above; one is tempted to speculate that they are equal, but we do not know of a proof and for now we take it to be a free parameter. 

This force will cause the vortices to drift; in a condition of steady state this must be balanced against the frictional force calculated above, leading to a velocity of 
\be
v = \frac{s_{tr}|\nabla T|}{\sigma} \ . 
\ee
As each vortex moves across the sample in the $x$ direction it causes a phase slip of $\frac{2\pi}{q}$, generating a voltage difference in the $y$ direction via the Josephson effect. Expressing the vortex density in terms of the magnetic field, we find the Nernst signal to be:
\be
e_N = \frac{3 B s_{tr}}{4\pi^3 \tilde{C}_D T^2}\,.
\ee
We anticipate further applications of this formalism. 

\section{Discussion} \label{sec:concl}
This has been a somewhat long journey, so we now summarize our results. We have presented an in-depth study of vortices in holographic superfluids and superconductors. We argued that the infrared physics can be usefully understood from the framework of defect CFT, which is elegantly geometrized by a $T=0$ near-horizon scaling solution that described a new kind of extremal black hole horizon: a Poincar\'e horizon with a bubble of Reissner-Nordstr\"{o}m horizon that carries a single unit of flux. We further solved the partial differential equations that captured the physics in the full UV geometry at finite temperatures, demonstrating that the low-temperature limits of various observables tended to the values obtained from the $T = 0$ scaling solution.

The embedding into the UV geometry allowed us to study the thermodynamics of vortices in detail. One novel result is that with superconducting boundary conditions, the thermodynamic stability of an $n = 2$ vortex as compared to two $n=1$ vortices can change, being unstable for small values of the bulk scalar charge $q$, but stable for larger values of $q$. This behavior is correlated with whether or not the superconductor is Type I or Type II, which should itself be reflected in the ratio of the London penetration depth $\lam$ and coherence length $\xi$ of the dual superconductor, an expectation that we confirm. Thus we conclude that holographic superconductors may be Type I over a range of parameters. 

Finally, we turned away from the gravitational description and discussed the forces on a moving conformal vortex. We demonstrated that there are simple expressions for these forces in terms of Kubo formulas of defect-localized operators. 

We note that some of the lessons from this analysis may be useful beyond the study of vortices. There has been a recent surge of activity in studying holographic systems with either explicit or spontaneous breaking of spacetime symmetries, by the addition of a lattice, vortices, stripes, etc. In many cases this 
symmetry breaking turns out to be irrelevant in the infrared. However sometimes -- for example in the current case, where the existence of a conserved magnetic flux guarantees that the inhomogeneity is transferred to the infrared -- this is not the case, and the low energy physics is strongly affected. We expect the notion of a conformal defect to be very useful in analyzing such situations and organizing the infrared behavior.

Our study suggests several directions for future research, which we discuss below:

\ben
\item One result from our analysis is that holographic superconductors {\it can} -- depending on the parameters -- be Type I. A more precise characterization of this property is possible. The key distinction between Type I and Type II superconductors is that the sign of the energy of the domain wall between the normal phase (with applied magnetic field) and the superconducting phase (with field expelled) is positive for Type I and negative for Type II. This can be studied directly in a holographic context by studying a different set of boundary conditions at the conformal boundary:
\begin{align}
F_{x_1 x_2}(x_1 \to -\infty) & = F_0, \qquad \langle \sO(x_1 \to -\infty) \rangle = 0\,, \\
F_{x_1 x_2}(x_1 \to +\infty) & = 0, \qquad \langle \sO(x_1 \to + \infty) \rangle = \langle \sO_0 \rangle \ . 
\end{align}
This will precisely create the domain wall in question, and its energy can now be studied explicitly. 

\item We have worked with zero chemical potential and induced our scalar field to condense at low temperature by adding a double trace deformation. It would be interesting to repeat our analysis of vortex stability for the more standard holographic superconductor, which starts with nonzero chemical potential and does not need a deformation.  It is not obvious that the results will be similar, since in the standard approach, increasing $q$ makes it easier for the scalar to condense, whereas here, increasing $q$ (in the presence of a magnetic field) makes it harder to condense.

\item In the regime where the superconductor is Type II, it is now natural to ask about a lattice of such vortices. A perturbative construction of such lattices has been initiated in \cite{Maeda:2009vf,Bao:2013fda,Bao:2013ixa}. While the explicit construction of such a lattice at zero temperature is difficult, armed with the results of this paper we may speculate about the ground state. From the scaling solution constructed in Section \ref{sec:confdef} we know that at zero temperature, even in the far infrared each vortex occupies a finite amount of proper cross-sectional area. Thus  for reasonably large lattice spacings we expect to simply have a regular lattice of vortices in the infrared, where each vortex is separated from the rest by regions of superconducting phase. While a single vortex preserves a near-horizon $SO(2,1) \times SO(2)$, this lattice will preserve only an infrared $SO(2,1)$. The construction of this near-horizon geometry remains an open problem. 

\item In our holographic model the vortex could be interpreted as a conformal defect because superfluid (or superconducting) order coexisted with a conformal sector down to arbitrarily low energies. It is of interest to understand whether this can happen in models with a more conventional UV description, e.g. a lattice Hamiltonian with short-range interactions. One such model where we do expect such a structure is in the $\mathbb{Z}_2$ fractionalized superconductor of \cite{PhysRevB.62.7850}. In that work two phases in $2+1$ dimensions called $SC$ and $SC^{\star}$ are described: both of them exhibit superconducting order but $SC^{\star}$ also supports a deconfined $\mathbb{Z}_2$ gauge field. Both phases are themselves gapped, containing a variety of heavy vortex and quasiparticle interactions. However they are separated by a continuous quantum phase transition in which the $\mathbb{Z}_2$ gauge field confines. This transition is in the 3d Ising universality class, and precisely at the critical point we expect that the vortex excitation in this model will flow to a conformal defect of the 3d Ising model. We expect our discussion in Section \ref{sec:confforce} to apply to this model, and in fact the properties of the Ising conformal defect in question have only recently been studied in \cite{Billo:2013jda}. It would be very interesting to understand other cases where this conformal vortex phenomenology could be applied. 

\item Finally, there is another way to interpret our results. Consider performing an S-duality in the bulk to re-interpret our calculation as the condensation of a {\it magnetically} charged scalar field. The bulk gauge field is now {\it confined} rather than Higgsed, and it is now electric flux that is confined to tight flux tubes, one of which we have constructed. It turns out that in the language of this paper the appropriate boundary conditions are those that we have labeled ``superconducting'': thus the bulk flux is allowed to penetrate the AdS boundary. Where each flux tube intersects the boundary it may now be interpreted as a heavy point charge. Thus the S-dual interpretation of our calculation is a state with a charge gap, or an {\it insulator} \cite{Faulkner:2012gt,Sachdev:2012tj}. Note that the quantization of electric charge is crucial to truly describe a phase as an insulator: in this S-dual construction this quantization arises from the fact that magnetic flux is quantized in each vortex. This is a novel kind of insulator, as electric charges are gapped, but  a neutral sector remains gapless. Our work in this paper amounts to the careful construction of a single gapped electric charge in this novel charge-gapped phase. 

This is a starting point towards an understanding of insulating phases, but much remains to be 
done. For example, it would be very interesting to try to construct a phase containing a finite {\it density} of flux, rather than a single unit of flux as studied here. In the Type I phase this would correspond after S-duality to a phase separated system containing macroscopic regions of ``insulator'' coexisting with phases of metal. In the Type II phase we would find a lattice of vortices, which would map to a Wigner crystal of charges. In this discussion we assume that we allow the charges to adjust their own spacing dynamically: if we instead impose a UV lattice periodicity by hand then we only expect to find an insulating phase when a commensurability condition is met, i.e. we require an integer number of quantized charges per lattice site. It would be quite interesting to understand such phases and the transitions to nearby metallic phases in more detail. 
\een

Clearly, there remain many open questions.  We can look forward to new insights as the holographic approach pursued here continues to illuminate  the physics of strongly correlated states of matter.

\begin{acknowledgments}
It is a pleasure to thank M.~Fisher, T.~Grover, S.~Hartnoll, K.~Jensen, S.~Sachdev and C.~H.~Yee for helpful discussions. This work was supported in part by the National Science Foundation under Grant No. PHY12-05500 and Grant No. PHY11-25915. J.E.S.'s work is partially supported by the John Templeton Foundation.

\end{acknowledgments}

\begin{appendix} 
\section{Boundary conditions along the axis - $x=0$:} \label{app:bc}

These boundary conditions are best understood if we first introduce the following radial variable:
\begin{equation}
R=\frac{x}{1-x}\,,
\end{equation}
in terms of which the line element (\ref{eq:ansatzmetric}) reduces to
\begin{multline}
\dd s^2 = \frac{L^2}{y^2}\Bigg\{-Q_1\,y_+^2 (1-y^3)\dd t^2+\frac{Q_2\,\dd y^2}{1-y^3}+\\
y_+^2Q_4\left[\dd R+\frac{y^2\,R\,Q_3}{(1+R)^2}\,\dd y\right]^2+y_+^2Q_5\,R^2\,\dd \varphi^2\Bigg\}.
\label{eq:ansatzmetricR}
\end{multline}

In addition, we also introduce cartesian coordinates
\begin{equation}
R=\sqrt{\tilde{x}^2_1+\tilde{x}^2_2},\quad\text{and}\quad \varphi = \arctan\left(\frac{\tilde{x}_2}{\tilde{x}_2}\right)\,.
\end{equation}

Recall that we want to ensure regularity at the axis $R=0$, i.e. that both the metric functions, scalar field and gauge field are regular in cartesian coordinates $(\tilde{x}_1,\tilde{x}_2)$. Let us first expand the line element (\ref{eq:ansatzmetricR}):
\begin{multline}
\dd s^2 = \frac{L^2}{y^2}\Bigg\{-Q_1\,y_+^2 (1-y^3)\dd t^2+\frac{Q_2\,\dd y^2}{1-y^3}+y_+^2(Q_4\dd x^2+Q_5\,R^2\,\dd \varphi^2)
\\+\frac{2\,y^2\,y_+^2\,Q_3\,Q_4(R\,\dd R)\,\dd y}{(1+R)^2}+\frac{R^2}{(1+R)^4}\,y^4\,y_+^2\,Q_4\,Q_3^2\,\dd y^2\Bigg\}\,.
\end{multline}
We can now read off the desired boundary conditions. First, the first term in the second line $R\,\dd R$ is a regular one form in cartesian coordinates, being equal to $\tilde{x}_1\dd \tilde{x}_1+\tilde{x}_2\dd \tilde{x}_2$. Second, the third term in the first line is only regular if  $Q_4=Q_5$. Under the above considerations, the above line element close to $R=0$ reduces to
\begin{multline}
\dd s^2 \approx \frac{L^2}{y^2}\Bigg\{-Q_1\,y_+^2 (1-y^3)\dd t^2+\frac{Q_2\,\dd y^2}{1-y^3}+y_+^2Q_4(\dd \tilde{x}^2_1+\dd \tilde{x}^2_2)
\\+2\,y^2\,y_+^2\,Q_3\,Q_4(\tilde{x}_1\,\dd \tilde{x}_1+\tilde{x}_2\,\dd \tilde{x}_2)\,\dd y+(\tilde{x}^2_1+\tilde{x}^2_2)\,y^4\,y_+^2\,Q_4\,Q_3^2\,\dd y^2\Bigg\}\,.
\end{multline}
The remaining boundary conditions are just found by noting that smooth functions of $(\tilde{x}_1,\tilde{x}_2)$, close to the cartesian origin, can only be functions of $\tilde{x}^2_1+\tilde{x}^2_2=R^2$, which translates into Neumann boundary conditions in $R$ for all the remaining \emph{metric functions}, \emph{complex scalar} and \emph{gauge field}. Finally, we need to rewrite these in terms of the original variable $x$. To summarize, the boundary conditions at the axis read
\begin{align}
&\left.\frac{\partial Q_1}{\partial x}\right|_{x=0}=\left.\frac{\partial Q_2}{\partial x}\right|_{x=0}=\left.\frac{\partial Q_4}{\partial x}\right|_{x=0}=\left.\frac{\partial Q_5}{\partial x}\right|_{x=0}=0\,,\quad Q_4(0,y)=Q_5(0,y)\,,\nonumber
\\
\\
&\left.\frac{\partial Q_3}{\partial x}\right|_{x=0}=2\,Q_3(0,y)\,,\quad\left.\frac{\partial Q_6}{\partial x}\right|_{x=0}=n\,Q_6(0,y)\quad \text{and}\quad\left.\frac{\partial Q_7}{\partial x}\right|_{x=0}=2\,Q_7(0,y)\,.\nonumber
\end{align}


\end{appendix}
\bibliography{refs}{}
\bibliographystyle{JHEP}

\end{document}